\documentclass[preprint,12pt,3p]{elsarticle}
\usepackage[pdfencoding=auto, psdextra]{hyperref}

\usepackage[section]{placeins}

\usepackage{amssymb}

\usepackage{subfig}

\usepackage{siunitx}

\usepackage{amsthm}

\usepackage{physics}

\usepackage{lineno}

\journal{Elsevier}

\begin{document}

\begin{frontmatter}

\title{The use of adversaries for optimal neural network training}

\author[label1]{Anton Hawthorne-Gonzalvez}
\address[label1]{School of Physics, The University of Melbourne, Parkville, Victoria, 3010, Australia}



\author[label1]{and Martin Sevior}


\begin{abstract}
B-decay data from the Belle experiment at the KEKB collider have a substantial background from $e^{+}e^{-}\to q \bar{q}$ events. To suppress this we employ deep neural network algorithms. These provide improved signal from background discrimination. However, the deep neural network develops a substantial correlation with the $\Delta E$  kinematic variable used to distinguish signal from background in the final fit due to its relationship with input variables. The effect of this correlation is reduced by deploying an adversarial neural network. Overall the adversarial deep neural network performs better than a Boosted Decision Tree algorithimn and a commercial package, NeuroBayes, which employs a neural net with a single hidden layer.  
\end{abstract}


\end{frontmatter}



\section{Outline}
\label{sec0}
This article is divided into sections as follows:
\begin{enumerate}
  \item {\bf Outline}
  \item {\bf The Physics Background (\ref{sec1})} describes the motivation for the particle physics measurement, the maain background, and the importance of the reducing its effect.
  \item {\bf Analysis of $B\to K_{s}\pi^{0}$ decays(\ref{sec2}}) We briefly describe the analysis procedure including event selection and the kinematic variabes employed for data analysis.
\item {\bf Kinematic Variables for Continuum Suppression (\ref{sec3})} which describes the variables employed by machine learning algorithimns to differentiate signal from background.
\item {\bf Deep Neural Networks with TensorFlow (\ref{sec4})} describes the deep neural net algorithim employed to distinguish signal from background. We also evaluate its performance and the observed correlation with the $\Delta E$ discriminating variable.
\item {\bf Adversarial Neural Networks With TensorFlow (\ref{sec5})} describes the development of the Adeversarial Neural Network algorithim (ANN) employed to reduce the correlation with $\Delta E$. We evaluate the performance of the ANN and find the optimum operating point.
\item {\bf Validation With Off-Resonance (\ref{sec6})} shows the performance of the machine learning algorithims with real data as opposed to the simulated data employed to develop the techniques.
\item {\bf Conclusions (\ref{sec7})} We summerize our work, show the performance of the technique along with its limitations and suggest an alternative approach for future study.
  
\end{enumerate}

\section{Physics Background}
\label{sec1}
    The Standard Model of particle physics (SM) is the most complete theory of the elementary particles and their interactions. It has substantial predictive power, yet there are compelling questions that cannot be answered by the SM. For example, what is the origin of Dark Matter \cite{Arkani-Hamed}? Why is there more matter than antimatter in the Universe \cite{Gavela}? These observations demand that there are fundamental new principles of nature beyond  that encompassed by the SM (New Physics). One way to learn the nature of New Physics is to search for discrepancies between measurements and SM calculations. To this end, rare decays of B-mesons, whose properties can be precisely predicted by the SM, are the subject of much experimental activity.
The Belle experiment employs the KEKB accelerator to collide $e^+e^-$ particles at the centre of mass energy of $\sqrt{s}=10.58$ GeV, corresponding to the $\Upsilon (4S)$ resonance. The $\Upsilon (4S)$ subsequently decays primarily (more than 96\% of the time) to $B\bar{B}$ pairs (48.6\% to $B^0\bar{B}^0$). Over the life of the experiment, $(771.581 \pm 10.566)\times10^6 B\bar{B}$ pairs were recorded and analysed with the detector. In addition, the KEKB collider also initiates the $e^{+}e^{-}\to q\bar{q}$ reaction where $q \in \{u,d,s,c\}$ (continuum background). This occurs at a rate 3 times greater than $\Upsilon (4S)$ production.
    
    Rare decays, such as the $B^0$ $\to K_S\pi^0$ decay, proceed via a $b\to u$ transition and are suppressed via the CKM matrix element V$_{ub}$ \cite{BtoKspi0_theory}. These typically have branching ratios of the order of $10^{-5}$ or smaller.  Therefore a large sample of $B\bar{B}$ pairs is required to make statistically significant measurements.

    
    \textit{CP}-symmetry is the expectation that applying $C$ (the charge operator - inverting all of the internal quantum numbers) and $P$ (the parity operator - reversing all spacial coordinates) would have no effect on the physics of a process.
    
    Both $B^0$ and $\bar{B}^0$ can decay to the $\textit{CP}$-eigenstate $K_S\pi^0$.  Direct \textit{CP}-violation (DCPV), is when these decay rates are not equal, and in a data sample with equal numbers of $B^0$ and $\bar{B}^0$ pairs, the DCPV is quantified with $\mathcal{A}_{CP}$ which is defined as:
    
    \begin{equation}
        \mathcal{A}_{CP}(K_S\pi^0) = \frac{N(\bar{B}^0\to K_S\pi^0) - N(B^0\to K_S\pi^0)}{N(\bar{B}^0\to K_S\pi^0) + N(B^0\to K_S\pi^0)}
    \end{equation}
    Where $N$ is the measured number of events for a given decay. The similar amplitudes of the two contributing Feynman diagrams mean that the process can exhibit relatively large DCPV.  The most recent Belle measurement is $\mathcal{A}_{CP}(K^0\pi^0)=+0.14\pm0.13(stat)\pm(0.06)(sys)$ \cite{Fujikawa:latestBelleAcp}, where the majority of the statistical uncertainty is due the large background to signal ratio. By combining measurements of all $K\to K\pi$ charge states, Beak et al. \cite{baek} predict that $\mathcal{A}_{CP}(K^0\pi^0)=-0.15\pm0.03$ and that a 5-$\sigma$ deviation from this value would demonstrate that the effects of New Physics make an unambiguous effect on this mode. Accordingly, the most precise measurements for  $\mathcal{A}_{CP}(K^0\pi^0)$ are vital. This is turn requires minimising the continuum background. We investigate advanced neural network (NN) machine learning algorithms to do this.
    
\section{Analysis of $B\to K_{s}\pi^{0}$ decays}
\label{sec2}

We begin by performing Monte-Carlo (MC) simulations of the $B\to K_S\pi^{0}$ process. We employ the EvtGen\cite{Ryd:2005zz} package in which the $\Upsilon(4S)$ decays to a $B^0\bar{B}^0$ pair. From here, one, $B^0_{sig}$, will decay to $K_S\pi^0$ and the other, $B^0_{tag}$, will decay generically. Particles from this process are propagated through the detector with Geant3\cite{Brun:1994aa}. Three data-sets of one-million events each are generated, for training, validation and testing of the neural networks. Continuum MC data is generated by the Belle collaboration at six times the expected yield (six streams) over the entire experiment, where two streams are used for NN training, one for validation, and three for testing.

    Real off-resonance (at a center of mass (COM) energy of $10.52\ \si{\GeV}$) data is available at 10.35\% of the expected continuum yield at the $\Upsilon(4S)$ resonance, and is used to finally validate NN performance.
    
    Two kinematic variables are employed in a maximum likelihood fit to discriminate between $B^0 \to K^0\pi^0$ events and backgrounds. These are $\Delta E$ and $M_{bc}^{corr}$. $\Delta E$ is the difference between the reconstructed $B$ meson energy ($E_B$) and half of the $e^+e^-$ COM energy, given by:
    
    \begin{equation}
        \Delta E = E_B - E_{beam}
    \end{equation}
    
    Where $E_{beam}$ is the beam energy in COM frame. $\Delta E$ peaks at 0.0 GeV for signal and has a continuous distribution for continuum background. Although $\Delta E$ peaks at 0.0 GeV for signal, it has a significant asymmetretic distribution due to energy leakage from the electromagnetic calorimeter (ECL), employed to measure the energy of the photons from the decay of $\pi^{0}$'s.
    $M_{bc}^{corr}$ is the beam constrained mass and is defined as:
    
    \begin{equation}
        M^{corr}_{bc} = \sqrt{E_{beam}^2-|\vec{p}_{B^0_{corrected}}|^2}
    \end{equation}
    
    Where $\vec{p}_{B^0_{corrected}}$ is the reconstructed $B$ meson momentum with a correction scale applied to the pion momentum to take account of ECL shower leakage. It is defined as:
    
    \begin{equation}
        \vec{p}_{B^0_{corrected}} = \vec{p}_{K_S} + \frac{\vec{p}_{\pi^0}}{|\vec{p}_{\pi^0}|}\sqrt{\left(E_{beam}-E_{K_S}\right)^2 - m^2_{\pi^0}}
    \end{equation}
    
    Where $\vec{p}_{K_S}$ and $E_{K_S}$ are the reconstructed $K_S$ momentum and energy respectively. $\vec{p}_{\pi^0}$ is the reconstructed pion momentum and $m_{\pi^0}$ is the $\pi^0$ world average mass. $ M^{corr}_{bc}$ peaks at the B-mass for signal and is continuously distributed up to $E_{beam}$ for continuum background.
    
    We place the selection criteria that $5.265\ \si{\GeV} c^{-2}<M_{bc}^{corr}<5.3\ \si{\GeV} c^{-2}$ and $-0.4\ \si{\GeV}<\Delta E<0.3\ \si{\GeV}$. This reconstruction and selection procedure leaves us with  $61385\pm143$ continuum events in one stream (this number is the number of events left in the sixth continuum stream, its uncertainty is its square-root as we assume a Poisson distribution). Of one-million signal MC events, 306,803 events remain, so the efficiency of reconstruction ($\epsilon_{recon}$) is $(30.68\pm0.06)$\%. From this we get the expected number of signal events:
    
    \begin{equation}
        N_{signal}=N_{B\bar{B}} \times R_{B^0\bar{B}^0} \times \mathcal{B}\left(B^0\to K^0\pi^0\right) \times \epsilon_{recon}
    \end{equation}
    Where $N_{B\bar{B}}$ (the total number of $B\bar{B}$ events) is $(771.581 \pm 10.566)\times10^6$, $R_{B^0\bar{B}^0}$ (the fraction of $B\bar{B}$ that are $B^0\bar{B}^0$) is $0.486\pm0.006$, $\mathcal{B}\left(B^0\to K^0\pi^0\right)$ is $(9.9\pm0.5)\times10^{-6}$\cite{Olive:2016xmw}. Note there is a factor of two (as there are two $B$ mesons that could decay in the signal channel) and a factor of 0.5 (the fraction of $K^0$ that go to $K_S$) that are not shown. This gives an expected signal yield of $1139\pm61$.

    There is far more continuum than signal (around 54 times as much) and it is vital to reduce this background before any physics analysis can proceed.

    \section{Kinematic Variables for Continuum Suppression}
    \label{sec3}
    In order to further reduce the continuum background, nineteen additional kinematic variables are calculated (taking advantage of the differing decay topologies between signal and continuum) and employed as input into a NN. The NN will then provide a classification (a variable upon which a selection criteria can be placed) based on these inputs.
    
    The nineteen variables employed by the NN are:
    \begin{itemize}
        \item $\Delta Z$: The distance along the beamline axis between the decay vertices of $B^0_{sig}$ and $B^0_{tag}$ in COM frame.
        
        \item $\operatorname{cos}(\theta_B)$: The angle between the reconstructed $B^0_{sig}$ momentum and the beamline in COM frame ($\theta_B$).
        
        \item $\operatorname{cos}(\theta_{thrust})$: $\operatorname{cos}$ of the angle between the signal thrust vector and the rest-of-event thrust vector. The thrust vector $\hat{n}$ is the unit vector which maximises the scalar thrust:
        \begin{equation}
            T = \frac{\sum_{i=1}^N |\hat{n}\cdot\vec{p}_i|}{\sum_{i=1}^N |\vec{p}_i|}
        \end{equation}
        Where $\vec{p}_i$ is the momentum of particle $i$.
        
        \item The sum of the transverse momenta (from the beamline) over all particles:
        \begin{equation}
            p^{sum}_t = \sum_{n=1}^{N} |\vec{p}_{t,n}|
        \end{equation}
        Where $\vec{p}_{t,n}$ is the transverse component of the momentum of particle $n$, and $N$ is the total number of particles. 
        
        \item The squared-missing-mass:
        \begin{equation}
            M_{miss}^2 = \left(2E_{beam}-\sum_{n=1}^{N}E_n\right)^2 - \left|\sum_{n=1}^{N} \vec{p}_n\right|^2
        \end{equation}
        Where $E_n$ and $\vec{p}_n$ are the energy and momentum of particle $n$ for all particles $n$.
        
      \item The Kakuno-Super-Fox-Wolfram moments \cite{kakuno} (KSFW): These essentially ``measure'' the degree to which the event shape is spherical. In the center of mass of the $e^{+}e^{-}\to B\bar{B}$ reaction, there is very little momentum delivered to the $B\bar{B}$ final state. In contrast the  $e^{+}e^{-}\to q\bar{q}$ continuum background has substantially more momentum delivered to the $q\bar{q}$ pair. After hadronising into mesons, the underlying momenta of the $q\bar{q}$ is preserved as predominantly back-to-back jets of mesons. In contrast, the decay of the $B\bar{B}$ results in a roughly spherical distribution of mesons since the velocities of the $B\bar{B}$ are small in the center of mass.
        The improved KSFW moments are divided into multiple categories, defined by the order ($l$) of the Legendre polynomials ($P_l$). They are further divided into `$oo$' and `$so$' when the sums are over just the rest-of-event particles or both rest-of-event particles and signal daughters. They are defined as follows:
        \begin{itemize}
            \item The `$so$' KSFW-moments of even order ($l=0,2,4$) are given by:
            \begin{equation}
            R^{so}_{xl} = \frac{\sum_{a}\sum_{b} |\vec{p}_b|P_l\left(\operatorname{cos}\left(\theta_{ab}\right)\right)}{E_{beam}-\Delta E}
            \end{equation}
            where $q$ is the particle charge, $a$ runs over the signal $B$ daughters and $b$ over the rest-of-event particles. This is further divided into three categories where $b$ runs only over charged ($x=0$), neutral ($x=1$) or missing ($x=2$) particles (i.e reconstructed momentum that doesn't correspond to measured particles). We then have nine KSFW moments of this type.
            
            \item The `$oo$' KSFW-moments for $l=1,3$ are given by:
            \begin{equation}
            R^{oo}_{l} = \frac{\sum_{a}\sum_{b} q_a q_b|\vec{p}_a||\vec{p}_b|P_l\left(\operatorname{cos}\left(\theta_{ab}\right)\right)}{(E_{beam}-\Delta E)^{2}}
            \end{equation}
            
            \item The `$oo$' KSFW-moments for $l=0,2,4$ are given by:
            \begin{equation}
            R^{oo}_{l} = \frac{\sum_{a}\sum_{b} |\vec{p}_a||\vec{p}_b|P_l\left(\operatorname{cos}\left(\theta_{ab}\right)\right)}{(E_{beam}-\Delta E)^{2}}
            \end{equation}
        
        \end{itemize}
    \end{itemize}

    \section{Deep Neural Networks with TensorFlow}
    \label{sec4}
    In order to maximise the continuum suppression, we build a NN architecture from the ground up using TensorFlow \cite{1603.04467:TensorFlow}. We  implement a deep NN and employ state of the art algorithms which together, have the potential to provide better classification than is achieved using single-layer NNs.
    
    The training data-set consists of 125,000 (correctly reconstructed) signal and 125,000 continuum events. First the data for each of the nineteen kinematic variables is pre-processed by implementing equal frequency binning. Here each variable is transformed so that the total 250,000 events, comprising signal and background, are evenly distributed among 500 equally spaced bins over the output range -1 to +1. During training, we employ the ADAM algorithm \cite{1412.6980:adamOptimiser} to minimise the cross-entropy ($\mathcal{L}_{class}$) as defined in equation \ref{classCrossEntropy}.
    
    \begin{equation}
    \label{classCrossEntropy}
        \mathcal{L}_{class}(\vec{x},\hat{y})=-\hat{y}\cdot log\left(y\left(\vec{x}\right) \right) - (1-\hat{y})\cdot log\left(1-y\left(\vec{x}\right)\right)
    \end{equation}
    
    Where $y(\vec{x})$ is the NN output given the vector of inputs (the kinematic variables) $\vec{x}$. $\hat{y}$ is the known (target) value (1 or 0 for signal or continuum respectively).
    
 The performance of the NN is measured by calculating the cross-entropy on the validation data-sets.
    The architecture of the NN (i.e. the number of hidden layers, nodes per layer and activation functions) and the training algorithm parameters (batch size, learning rate, training steps etc.) are the hyper-parameters that must be specified. The hyper-parameter space is immense so finding the best configurations requires an efficient algorithm. The best hyper-parameter configuration of the TensorFlow network was found using HyperBand \cite{1603.06560:HyperBand}, which narrows down the best configuration from a large random sample of hyper-parameter configurations. It achieves this by training a few configurations over the full training run, many configurations for a small fraction of the full training run, and a range in between. This combines the need to check many configurations with the need to better evaluate each given configuration. Once the best configuration is found, the NN is trained with this set of hyper-parameters. The NN is then applied to the testing data-set and employed for further physics analysis. 
    
    The hyper-parameter configuration for the best performing network is as follows:
        
    \begin{itemize}
    \setlength\itemsep{0em}
        \item A maximum number of epochs (the number of times the training runs of the entire data) of 600.
        \item 50 events per batch (the number of events in each training step).
        \item A Learning rate of 0.0001.
        \item Six hidden layers.
        \item 47 nodes per hidden layer.
        \item Exponential linear unit activation function.
    \end{itemize}
    
\begin{figure}[h!]
      \begin{center}
        \includegraphics[width=0.8 \columnwidth,height=!,angle=0]{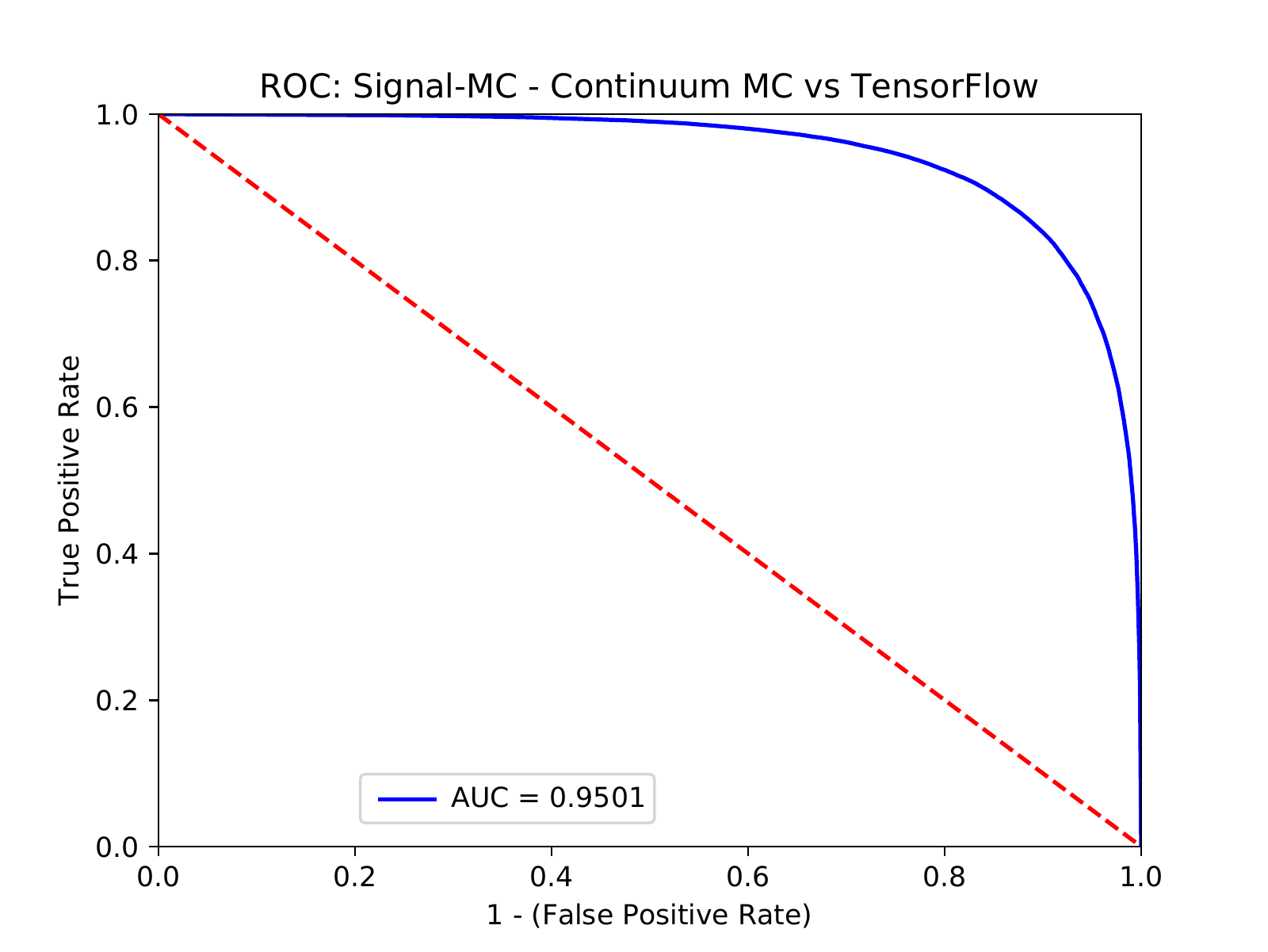}
      \end{center}
      \caption
      %
          {The Receiver - Operator Curve for the trained TensorFlow network (TF1) as applied to the the continuum and signal testing datasets. Only loose cuts on $\Delta E$ of $-0.4 \:GeV < \Delta E < 0.3 \: GeV$ are applied. The AUC = 0.9501. The broken red line shows no training.}
      \label{figChapt7:TFNNOutputSigContEqualNumbers}
    \end{figure}

    The Receiver-Operator Curve (ROC) ($\mathit{NN}$) of the optimised TensorFlow NN (TF1), is shown in figure \ref{figChapt7:TFNNOutputSigContEqualNumbers}. This and the following two figures also show the Area Under the Curve (AUC). The AUC quantifies the overall effectiveness of the classification. An AUC = 0.5 implies no effect while AUC = 1.0 implies perfect classification. These results were obtained from processing the testing data sets.
    
    \subsection{Analysis of the TensorFlow Neural-Network Performance}
    \label{subsec:TFNeuralNetAnalysis}
    
    To evaluate TF1, we compare it to the performance of the NeuroBayes neural network package (NB) \cite{Feindt:NeuroBayesPaper} as well as the Boosted Decision Tree (BDT) algorithim as implmented by the TMVA - Toolkit for MultiVariate Analysis package \cite{TMVA}. Both NB and the BDT method of TMVA are widely used by the Belle Collaboration for event classification.
    
    The internal architecture of NB consists of one hidden layer where the number of nodes was set to the default value of 21.

    The input data is pre-processed by NB to transform each variable into a Gaussian distribution. The batch-size was 100, and NB was trained over 150 epochs using the Broyden–Fletcher–Goldfarb–Shanno algorithm (see \cite{Byrd1995:BFGSAlgorithm} for more information). Regularisation is employed using the `Bayesian regularisation procedure' (see \cite{Feindt:NeuroBayesAlgorith} for details on the NeuroBayes algorithm). During training NeuroBayes employs pruning and removal of the least important weights to prevent over-training.

\begin{figure}[h!]
      \begin{center}
        \includegraphics[width=0.8 \columnwidth,height=!,angle=0]{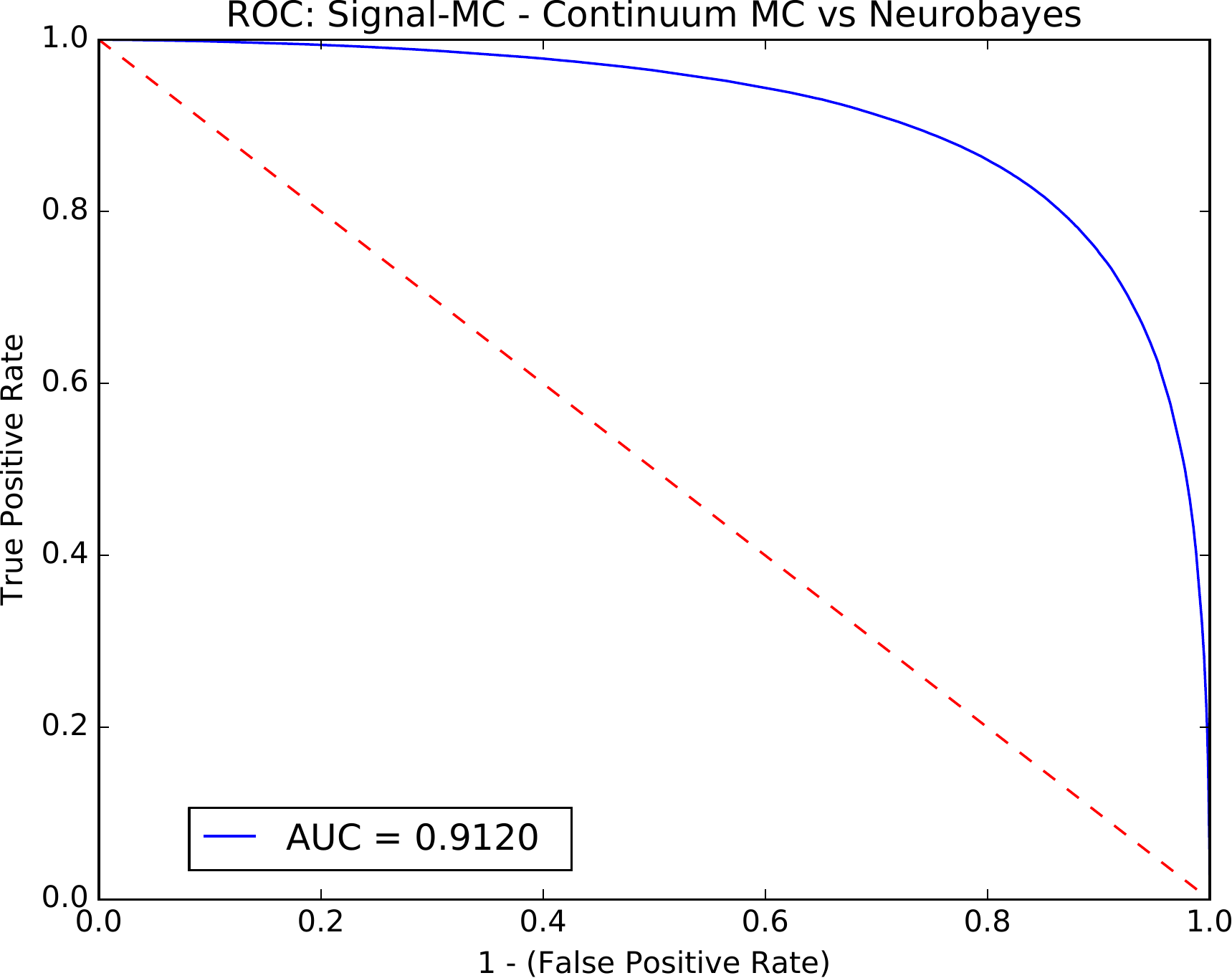}
      \end{center}
      \caption
      {The Receiver - Operator Curve (ROC) for the NeuroBayes neural network (NB) output, $\mathit{NN}$ for the continuum and signal testing datasets. Only the loose cuts on $\Delta E$ are applied. The AUC = 0.912. The broken red line shows no training.}
      \label{figChapt6:NNoutputEqualNumbers}
    \end{figure}

    NB was trained on the same data-sets as TF1. The set up was not tweaked to improve performance, therefore this network was not applied to the validation data-sets, and instead applied directly to the testing data-sets. The  Receiver - Operator Curve (ROC) output of the trained NeuroBayes NN can be seen in Figure \ref{figChapt6:NNoutputEqualNumbers}.
    
\begin{figure}[h!]
      \begin{center}
        \includegraphics[width=0.95 \columnwidth,height=!,angle=0]{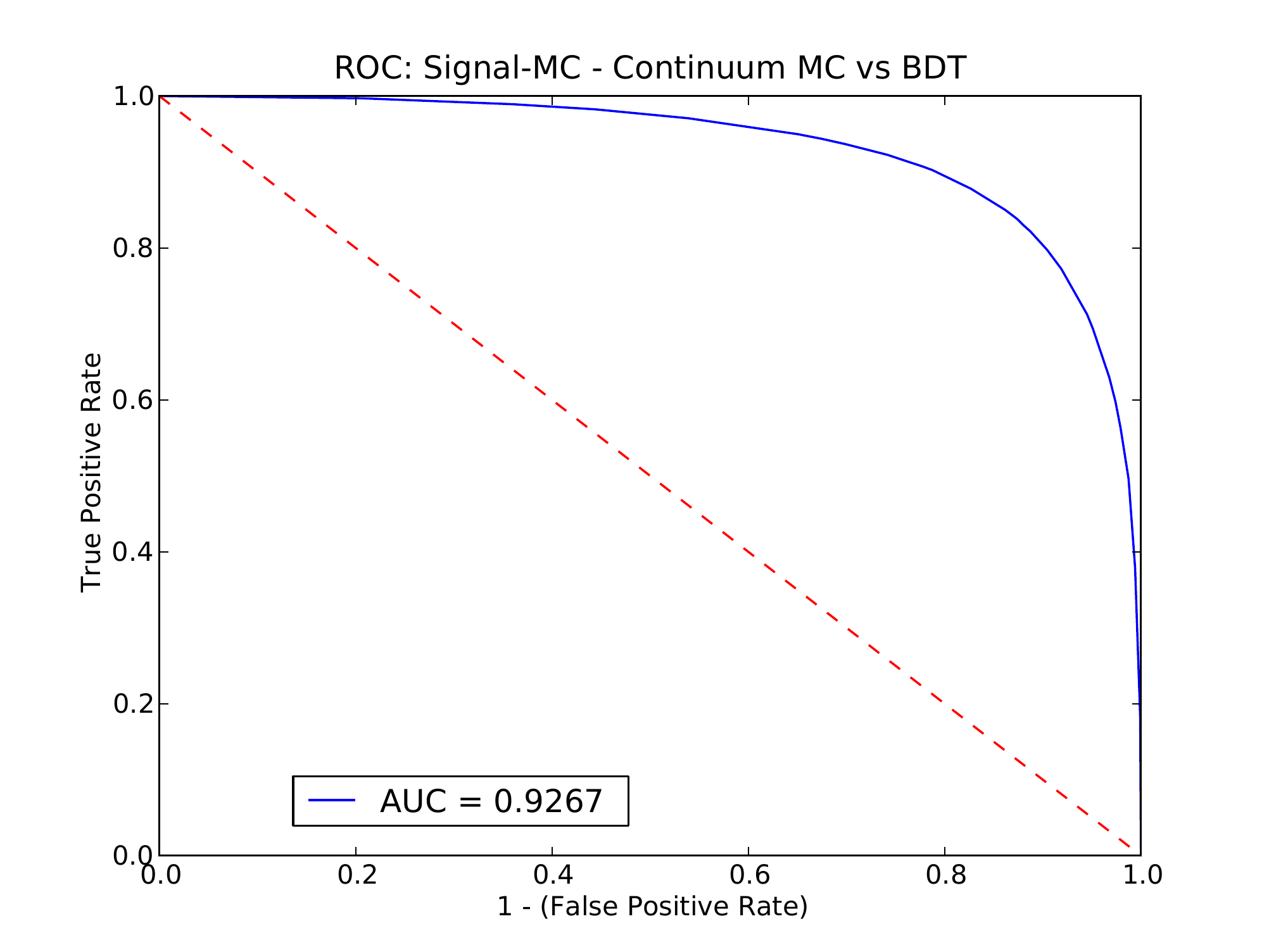}
      \end{center}
      \caption
      {The Receiver - Operator Curve (ROC) of the TMVA BDT for the continuum and signal testing datasets. Only the loose cuts on $\Delta E$ are applied. The AUC = 0.9267. The broken red line shows no training.}
      \label{BDT:ROC_nodele}
    \end{figure}

    The performance of the TMVA BDT is shown in figure \ref{BDT:ROC_nodele} where the performance was measured using the same testing data sets as NB. The AUC of the algorithim is 0.9267.

    For TF1, a value of $\mathit{NN}_{cut}$ (a selection criteria for which $\mathit{NN}>\mathit{NN}_{cut}$) chosen to keep 13.00\% of continuum, leaves 88.11\% of signal (in contrast with 79.01\% from NB). Alternatively a value of $\mathit{NN}_{cut}$ chosen to keep 70.20\% of signal using TF1, leaves 3.35\% of continuum remaining (in contrast with 7.65\% from NB). These results show that, depending on the $\mathit{NN}_{cut}$ choice, the continuum background for a given signal efficiency could be reduced by over a factor of 2 by employing TF1 rather than NB.

    \subsection{$\Delta E$ - Classifier Correlations}

\begin{figure}[h!]
          \begin{center}
            \subfloat{
            \includegraphics[width=0.95 \columnwidth,height=!,angle=0]{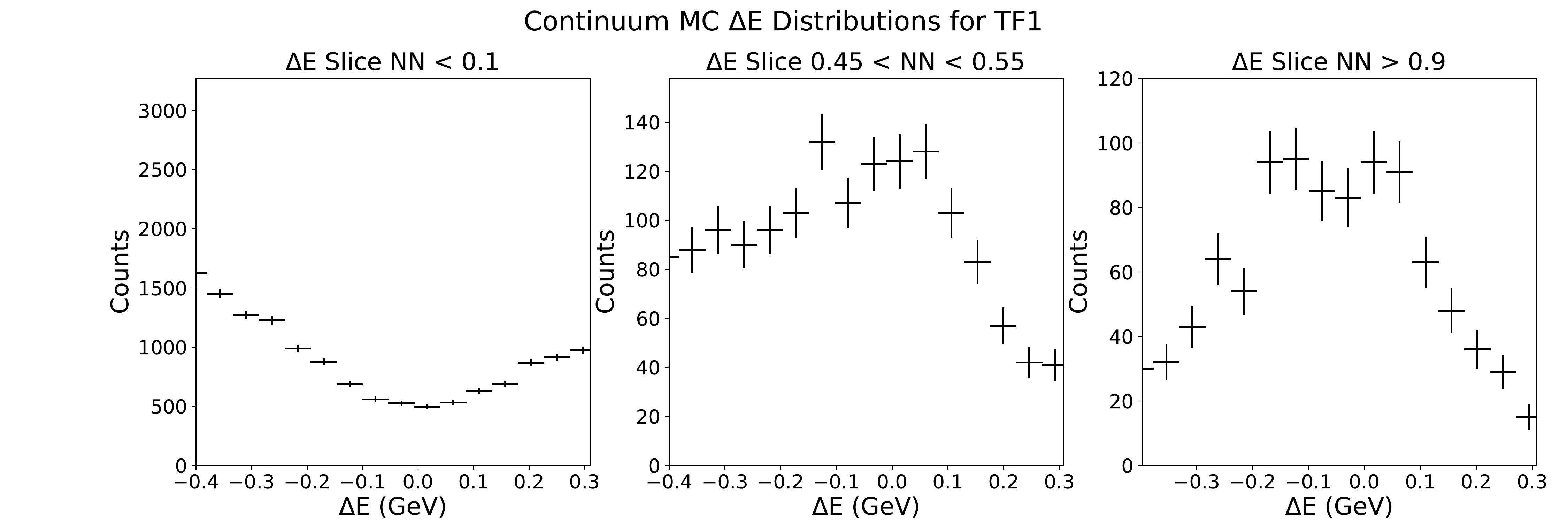}
            } \\

            \subfloat{
            \includegraphics[width=0.95 \columnwidth,height=!,angle=0]{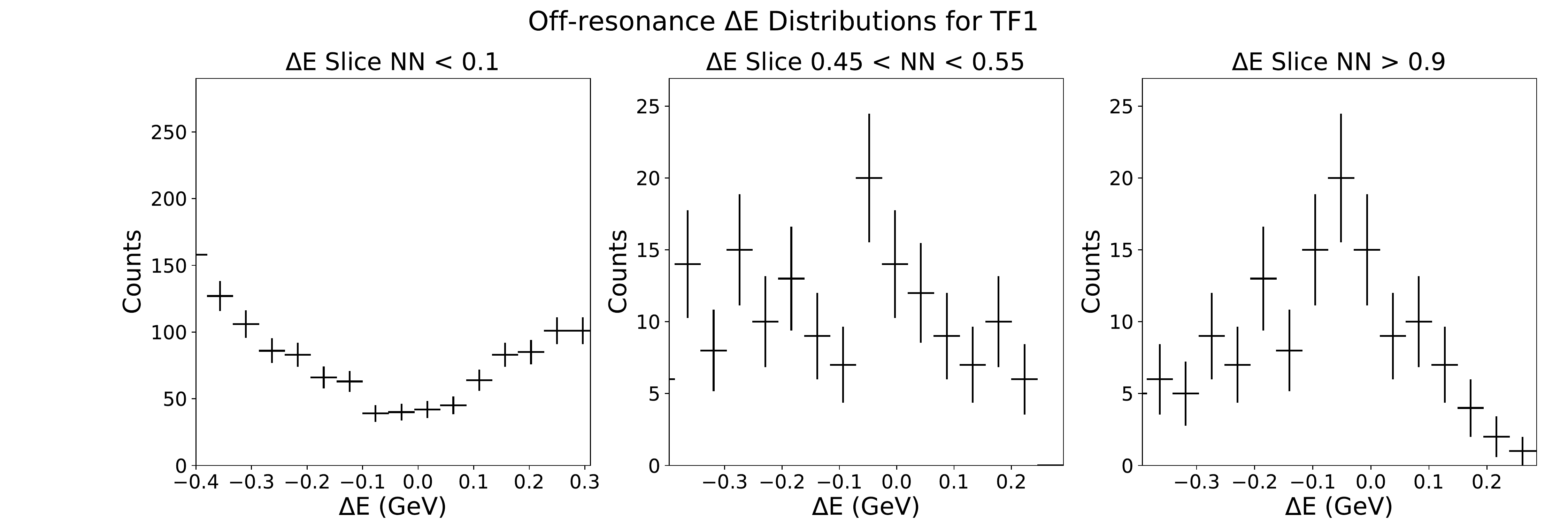}
            }

          \end{center}
          \caption
          {Showing the scultping of the $\Delta E$ distributions at different $\mathit{NN}$ (TF1) slices for $e^{+}e^{-}\to q \bar{q}$ events. The effect is seen in both continuum MC (top row) and off-resonance real data (bottom row). }
          \label{figChapt7:DeleTFNNSliceContinuum}
        \end{figure}

\begin{figure}[h!]
          \begin{center}
            \subfloat{
            \includegraphics[width=0.95 \columnwidth,height=!,angle=0]{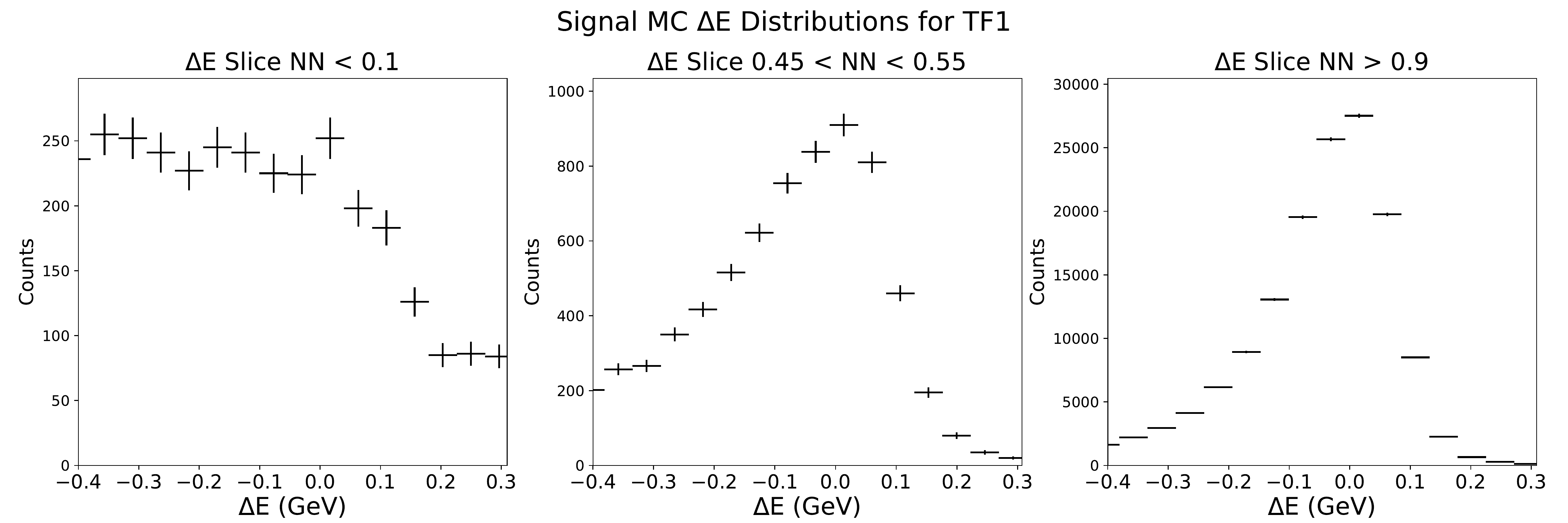}
            }
          \end{center}
          \caption
          {Showing the signal $\Delta E$ distributions for different $\mathit{NN}$ (TF1) slices. }
          \label{figChapt7:DeleTFNNSliceSignal}
        \end{figure}

\begin{figure}[h!]
          \begin{center}
            \subfloat{
            \includegraphics[width=0.3 \columnwidth,height=!,angle=0]{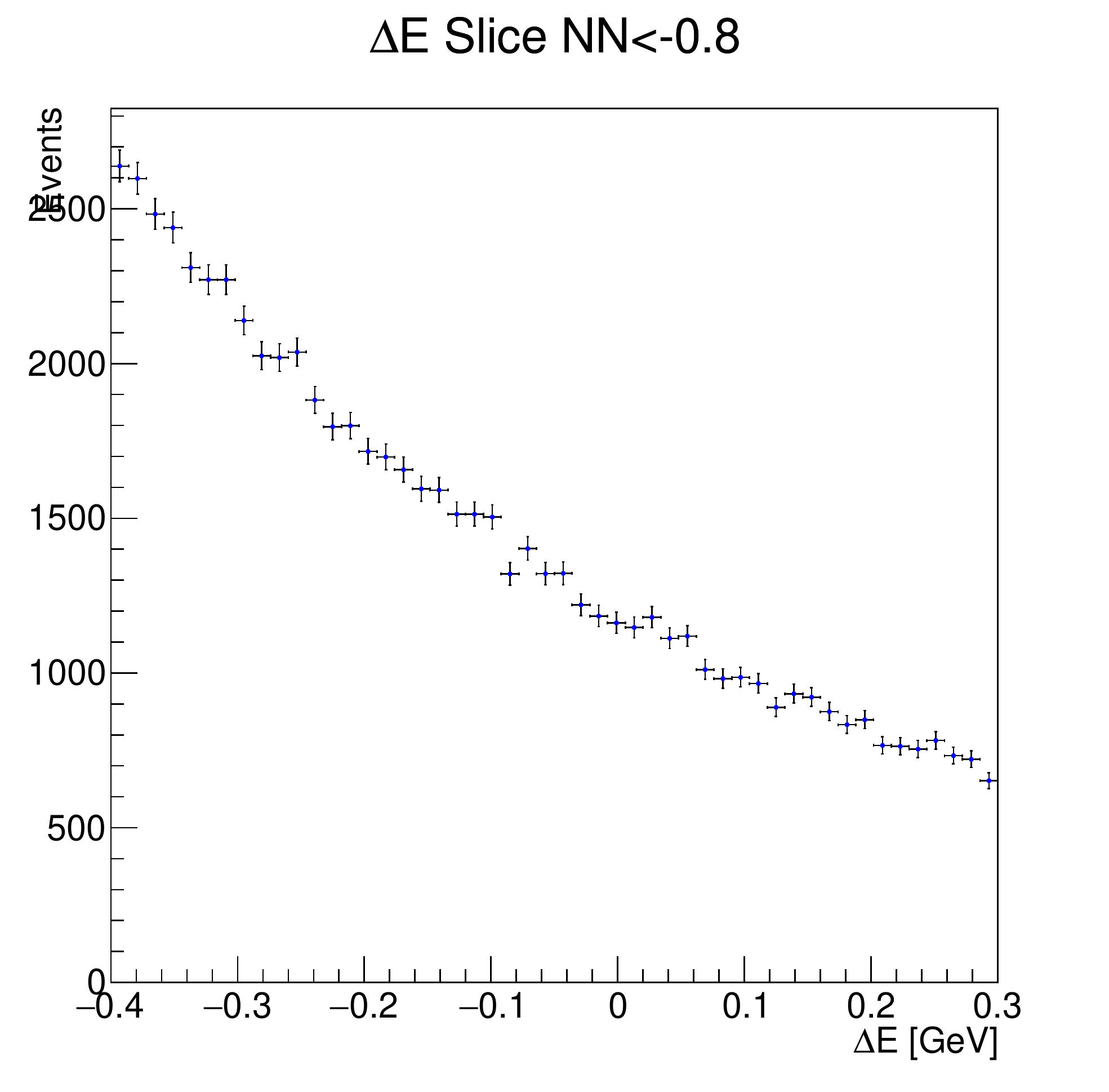}
            }
            \subfloat{
            \includegraphics[width=0.3 \columnwidth,height=!,angle=0]{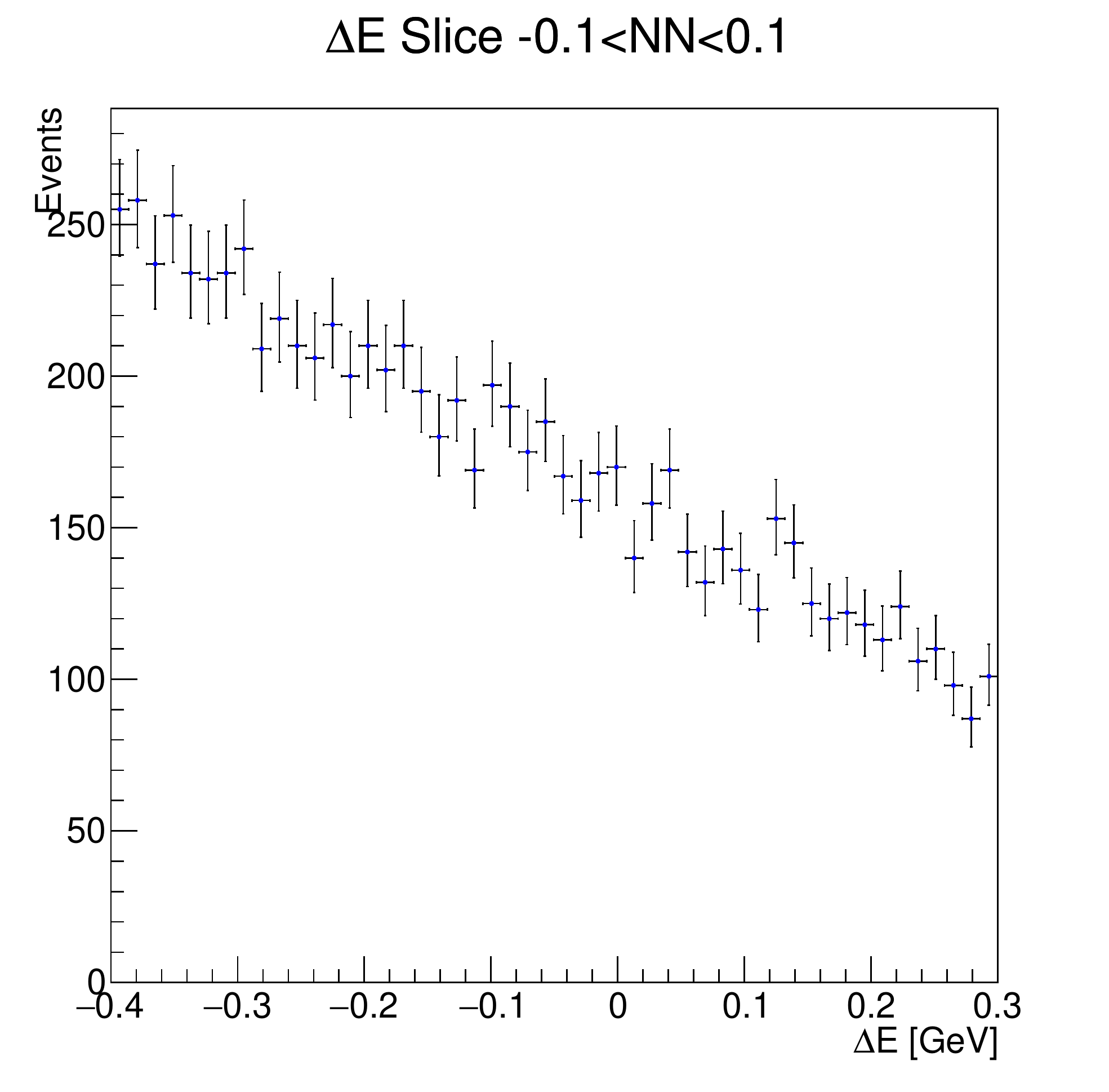}
            }
            \subfloat{
            \includegraphics[width=0.3 \columnwidth,height=!,angle=0]{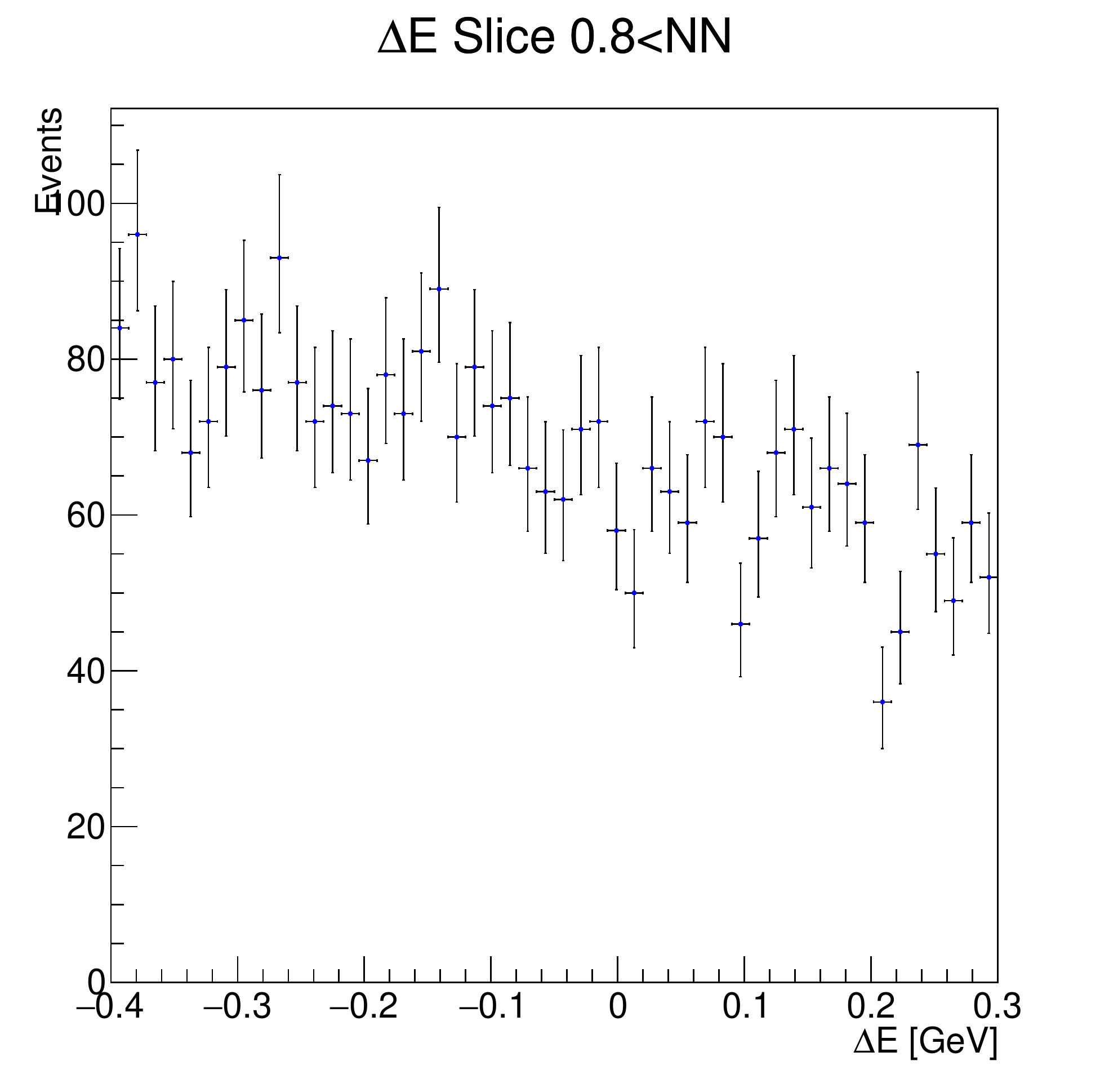}
            }
            \\
            \subfloat{
            \includegraphics[width=0.3 \columnwidth,height=!,angle=0]{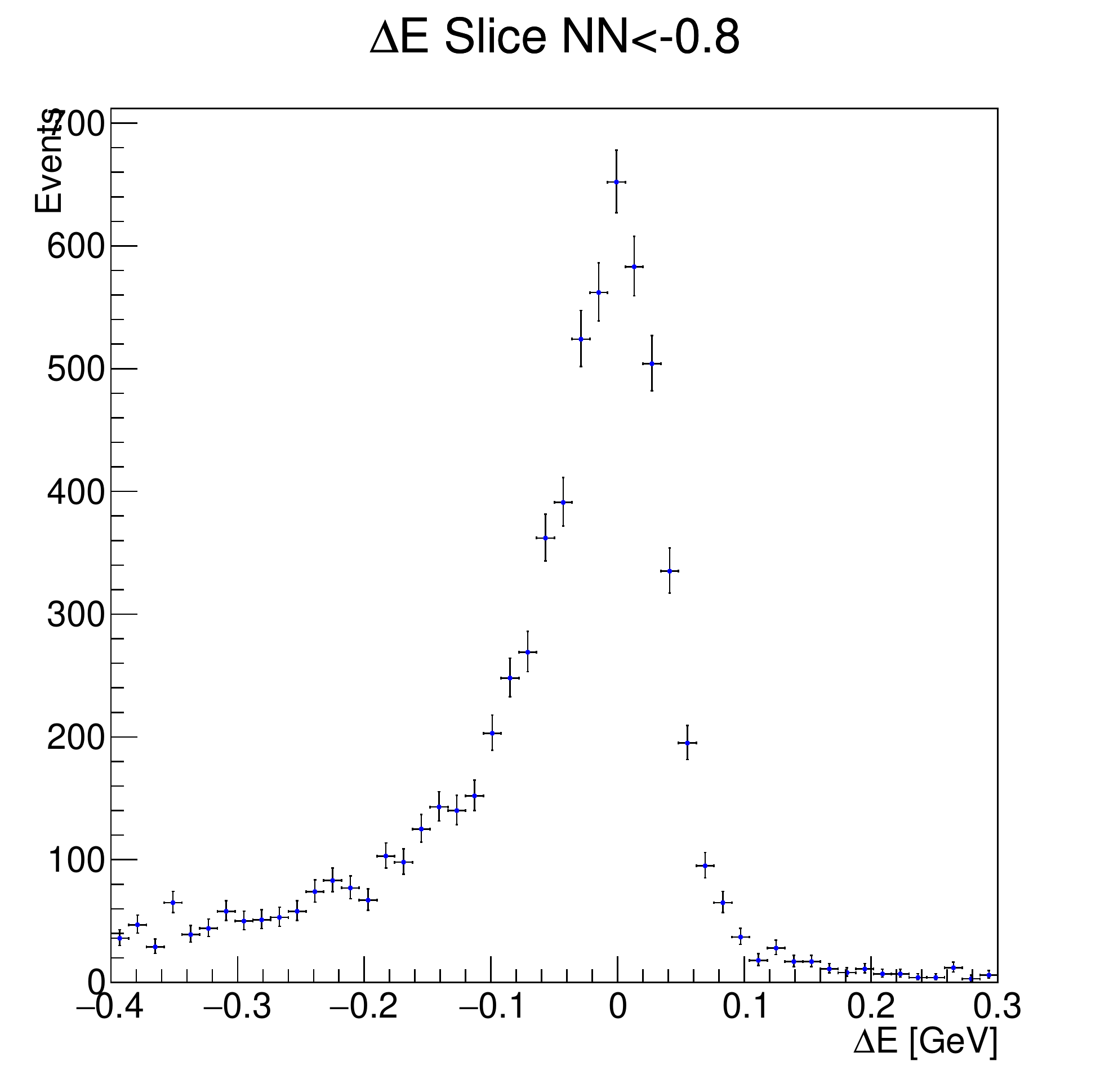}
            }
            \subfloat{
            \includegraphics[width=0.3 \columnwidth,height=!,angle=0]{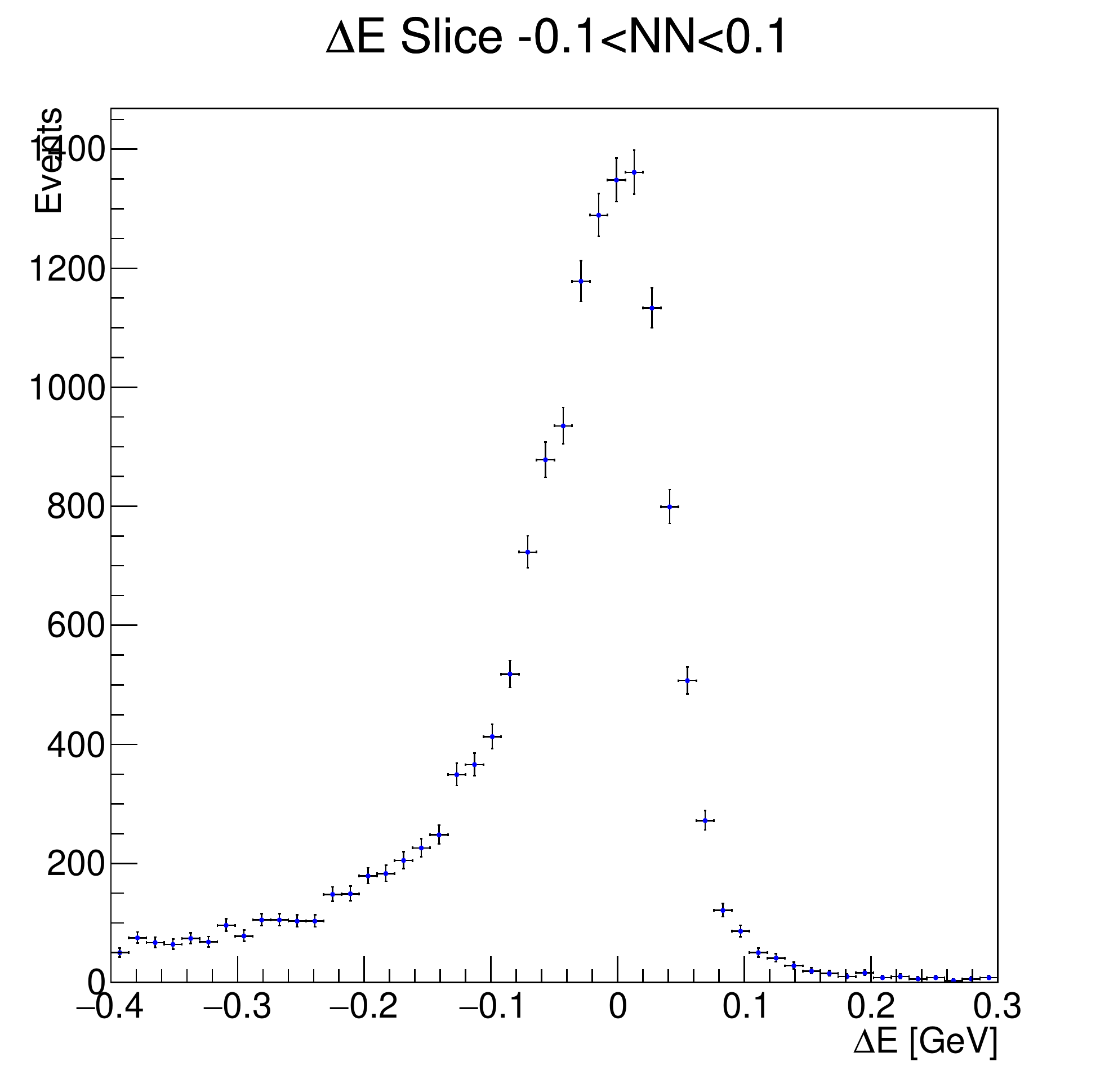}
            }
            \subfloat{
            \includegraphics[width=0.3 \columnwidth,height=!,angle=0]{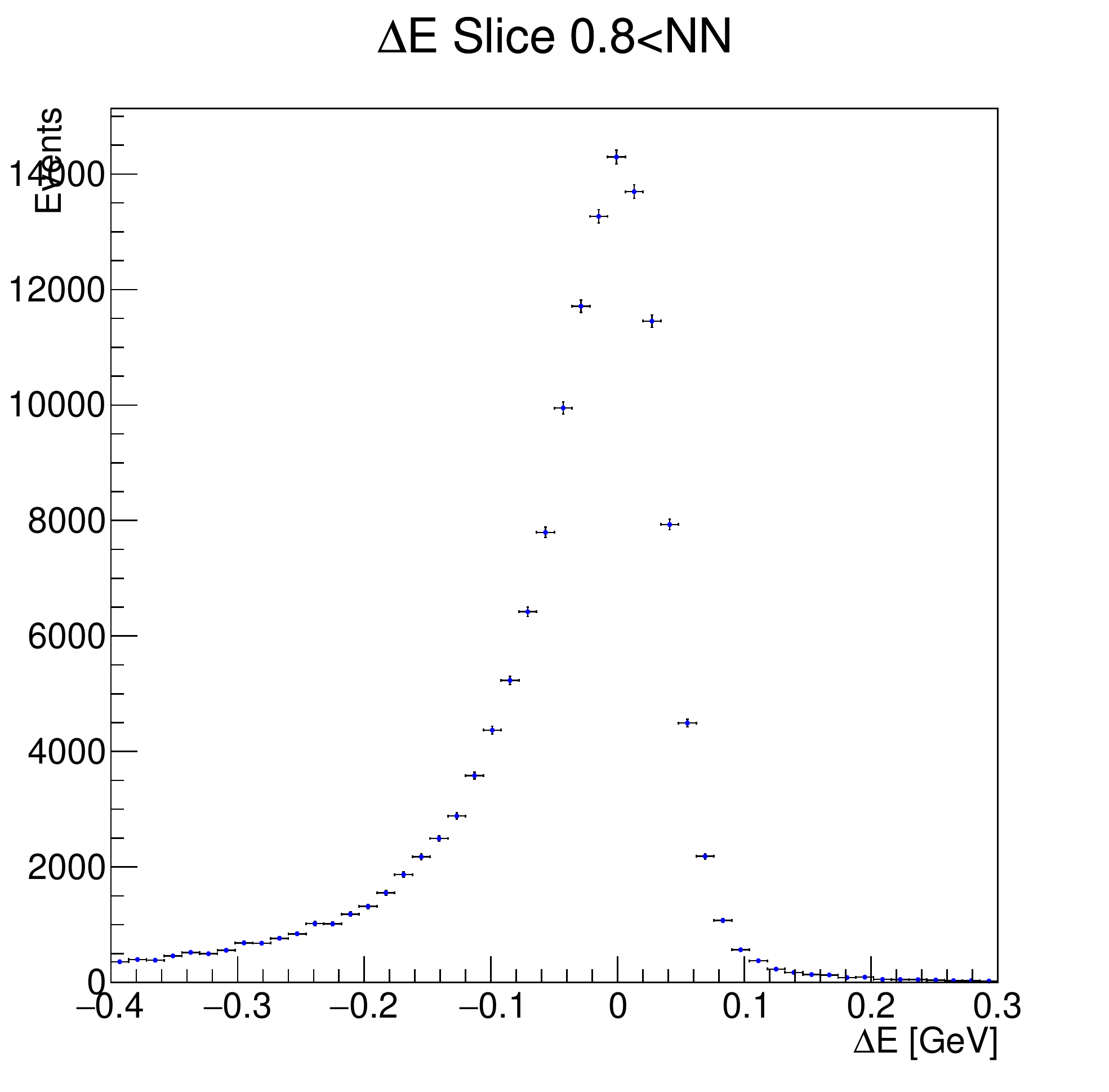}
            }
          \end{center}
          \caption
          {Showing the continuum (top row) and signal (bottom row) $\Delta E$ distributions at different slices of $\mathit{NN}$ from the NeuroBayes NN (NB). Note that as $\mathit{NN}$ is in the range $\pm1$ for the NeuroBayes output, the $\mathit{NN}$ slices are adjusted accordingly. }
          \label{figChapt7:NeuroBayesDeleNNSlicesSignalContinuum}
        \end{figure}

\begin{figure}[h!]
          \begin{center}
            \subfloat{
            \includegraphics[width = 1.0 \columnwidth]{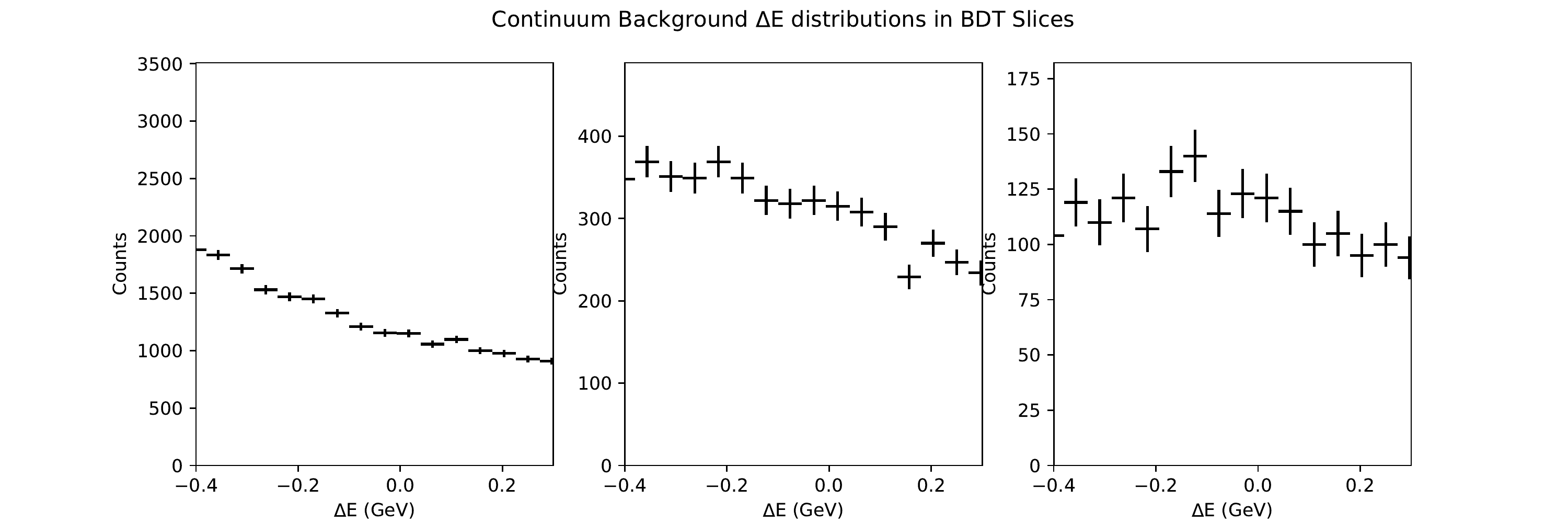}
            }
            \\
            \subfloat{
            \includegraphics[width = 1.0 \columnwidth]{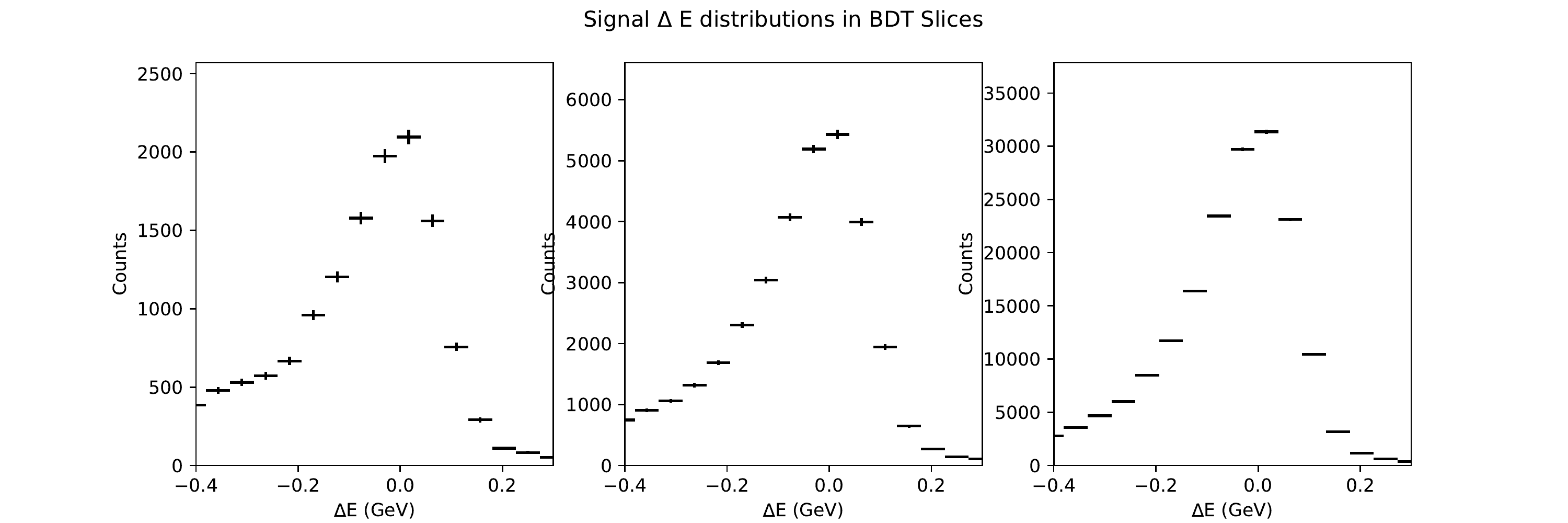}
            }
          \end{center}
          \caption
          {Showing the continuum (top row) and signal (bottom row) $\Delta E$ distributions at different slices of the classification variable (BDT) from the Boosted Decision Tree algorithim. The range of the slices of the BDT approximately matches those of the $\mathit{NN}$'s. The ``low slice'' is predominantly background, the ``high slice'' is predominantly signal.}
          \label{BDTDeleSlicesSignalContinuum}
        \end{figure}

        Investigation into the $\Delta E$ distribution at different $\mathit{NN}_{cut}$ values for TF1 shows unexpected results. When looking at the continuum $\Delta E$ distributions for different $\mathit{NN}$ slices, the distribution is sculpted to be more signal-like as $\mathit{NN}$ increases. On the flip side, for a low $\mathit{NN}$ the distribution shows the reverse, a trough where signal peaks. Figure \ref{figChapt7:DeleTFNNSliceContinuum} shows the continuum $\Delta E$ distributions at different $\mathit{NN}$ ranges. The off-resonance data also shows the same effect. Similarly for the signal distribution (see Figure \ref{figChapt7:DeleTFNNSliceSignal}), where the $\Delta E$ distribution becomes less signal-like as $\mathit{NN}$ decreases. This effect was not evident when using NB as  demonstrated in figure \ref{figChapt7:NeuroBayesDeleNNSlicesSignalContinuum}. This shows the $\Delta E$ distributions for different NeuroBayes NN output slices, for signal and continuum respectively. Nor is it evident when using the BDT algorithim. Figure \ref{BDTDeleSlicesSignalContinuum} shows the $\Delta E$ distributions for different BDT output slices over approximately the same classification ranges as the neural nets.

        This correlation can also be observed by looking at the continuum rejection percentage in an enhanced signal region of $-0.1<\Delta E<0.1$ (``tight region"). A $\mathit{NN}_{cut}$ selection is imposed on both NN outputs to keep the signal efficiency the same; $92.5\%$ over the full $\Delta E$ range. The continuum rejection rate for the TensorFlow NN is $80.09\%$ over the full $\Delta E$ range but only $63.4\%$ in the tighter region. In comparison, the NeuroBayes NN, has continuum rejection rates of $66.7\%$ and $64.08\%$ for the full and tighter $\Delta E$ regions.
        
        The TensorFlow NN learned that there is a relation between the $\Delta E$ value, and whether an event is signal or continuum. This is not observed by either the NB or the BDT algorithimns. The behaviour of TF1 is due to correlations between $\Delta E$ and the kinematic variables on which the NN is trained.  The kinematic variables with the largest correlations were found to be, in decreasing order: $R^{so}_{20}$, $R^{oo}_{0}$, $R^{oo}_{2}$ and $R^{so}_{22}$, with signal MC(continuum MC) correlations of 29.1\%(43.0\%), 18.2\%(27.1\%), 12.6\%(19.8\%), and 13.4\%(17.0\%) respectively.

        \section{Adversarial Neural Networks With TensorFlow}
        \label{sec5}
    The correlation with $\Delta E$ significantly reduces the effectiveness of the continuum suppression of TF1 in the signal region when performing a unbinned maximum likelihood fit to the data. While removing $R^{so}_{20}$, $R^{oo}_{0}$, $R^{oo}_{2}$ and $R^{so}_{22}$ would reduce the effect, there is discriminating power in the variables which we would lose if we adopt this approach. An alternative was to investigate employing an adversarial neural network (ANN).
    
    ANNs are used in order to generate images from a trained image-recognition convolutional NN, and used to further train the convolutional NN in order to perform better in classification tasks \cite{1406.2661:GenerativeAdversarialNetwork}. The idea of an ANN can be used to reduce the correlations between the output of a NN (referred to as the classifying neural network) and other parameters associated with the event. This method is used to reduce the correlation between $\mathit{NN}$ and $\Delta E$, although in principle this could also be used for any one of, or multiple parameters that have correlations with $\mathit{NN}$. The method laid out here closely follows that in \cite{1611.01046:AdversarialNetworkPivot}.

    The adversarial network models the $\Delta E$ distribution by taking $\mathit{NN}$ as input, and predicting the value of $\Delta E$ that a given event will have. The adversarial network used in this study has one input ($\mathit{NN}$), two hidden layers with 20 nodes each, and 15 outputs. The result is to model $\Delta E$ with five Gaussians (indexed by $i$), with 3 outputs for each, corresponding to the means ($\mu_i(\mathit{NN})$), widths ($\sigma_i(\mathit{NN})$), and fractional weighting of that Gaussian ($f_i(\mathit{NN})$). The fractions are not normalised so they are first passed through a softmax function (giving $f'_i(\mathit{NN})$, scaled to sum to one). The adversary loss function (for a single event) is given by:

    \begin{equation}
    \label{chapt8equation:adversaryLoss}
        \mathcal{L}_{adv}\left(\mathit{NN},\Delta E\right) = -log\left(\sum_{i=1}^5\frac{f'_i\left(\mathit{NN}\right)}{\sqrt{2\pi\sigma_i^2\left(\mathit{NN}\right)}}\exp{\frac{-\left(\mu_i\left(\mathit{NN}\right)-\Delta E\right)^2}{2\sigma_i^2\left(\mathit{NN}\right)}}\right)
    \end{equation}

    Training the ANN to minimise this loss function allows it to predict the $\Delta E$ distribution from the input $\mathit{NN}$ distribution if the two parameters are correlated and the $\Delta E$ distribution can be modelled with five Gaussians. We want to penalise the classifying neural network if $\mathit{NN}$ is correlated to $\Delta E$, so the classifier is further trained to minimise:

    \begin{equation}
    \label{chapt8equation:totalLoss}
        \mathcal{L}_{tot} = \mathcal{L}_{class} - \lambda_{adv}\mathcal{L}_{adv}
    \end{equation}

\begin{figure}[h!]
      \begin{center}
        \includegraphics[width=0.8 \columnwidth,height=!,angle=0]{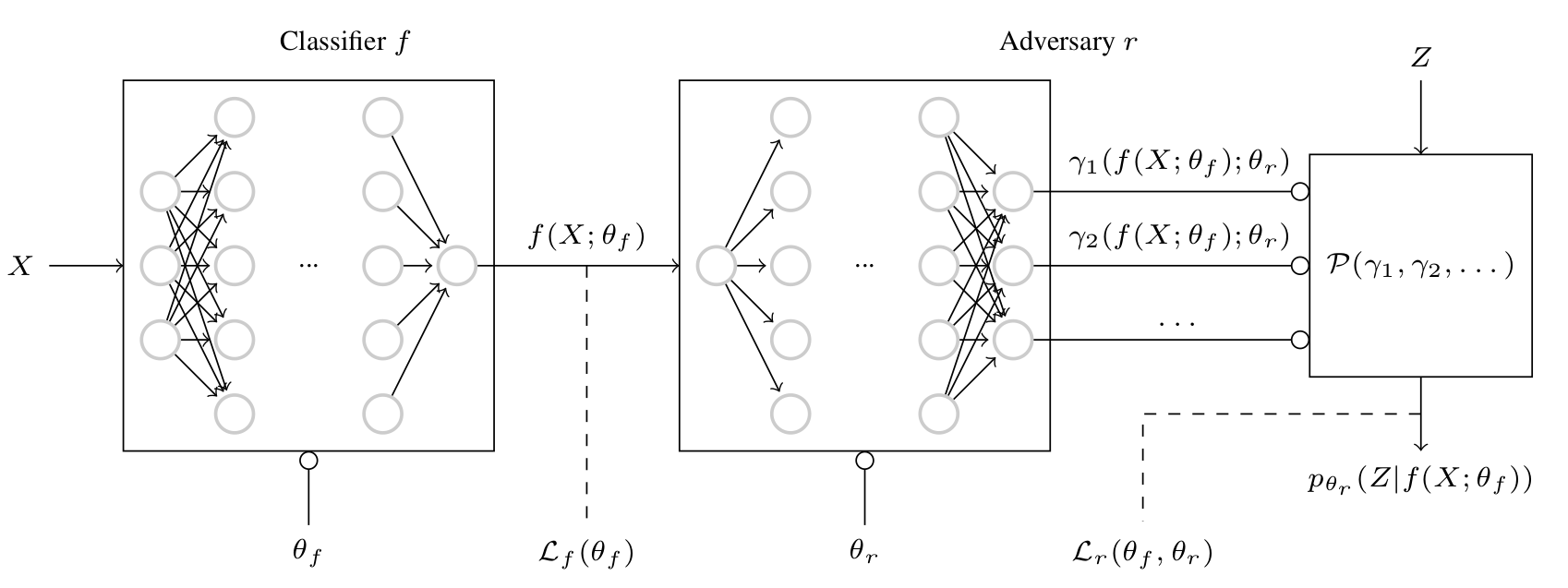}
      \end{center}
      \caption
      {Showing the configuration of the classifying and adversarial neural networks. $\theta_f$ and $\theta_r$ are the trainable-weights in the classifier and adversarial network respectively. $X$ is the vector of input kinematic variables. $f(X;\theta_f)$ is $\mathit{NN}$. $Z$ is $\Delta E$. $\gamma_{1-15}$ are the Gaussian means, standard-deviations and fractions, and $\mathcal{P}$ is the function that combines these (with $\Delta E$) into the likelihood function $p_{\theta_r}$. $\mathcal{L}_f(\theta_f)$ and $\mathcal{L}_r(\theta_f,\theta_r)$ are $\mathcal{L}_{class}$ and $\mathcal{L}_{adv}$ respectively. Image from \cite{1611.01046:AdversarialNetworkPivot}.}
      \label{figChapt8:AdversarialNetDiagram}
    \end{figure}

    Where $\mathcal{L}_{class}$ is our loss function for the classifier network (the cross entropy defined in \ref{classCrossEntropy}), and $\lambda_{adv}$ is a constant chosen to specify how much to penalise $\Delta E - \mathit{NN}$ correlations. Training to this new loss function has the desired impact of reducing the correlations at the cost of the continuum suppression. A $\lambda_{adv}$ of zero would result in a classifier identical to TF1, whereas a larger $\lambda_{adv}$ results in reduced $\mathit{NN}-\Delta E$ correlations but worse classifying power over the whole range of $\Delta E$. The configuration of the networks is shown in Figure \ref{figChapt8:AdversarialNetDiagram}.

    There are now the additional hyper-parameters associated with the architecture of the adversary network. These include the number of outputs (related to the number of Gaussians with which to model the $\Delta E$ distribution), the number of hidden layers and nodes per hidden layer. The additional hyper-parameters associated with training the adversary network are its batch-size, training steps and learning rate.

    The classifier neural network has the same architecture, and is initialised to the optimal weight values from the previous training in \ref{subsec:TFNeuralNetAnalysis}. For every classifier training step, the adversary network is first trained for 100 steps using the ADAM optimiser with a batch size of 125 and a learning rate of 0.01. The learning rate for training the classifier is reduced to $10^{-6}$ and the training is run for 4 epochs.

    The main hyper-parameters for the classifier are:
    \begin{itemize}
    \setlength\itemsep{0em}
    \footnotesize{
            \item A maximum number of epochs of 4.
            \item 50 events per batch.
            \item A Learning rate of $10^{-6}$.
            \item Six hidden layers.
            \item 47 nodes per hidden layer.
            \item Exponential linear unit activation function.
    }
    \end{itemize}

    And the hyper-parameters associated with the ANN are:
    \begin{itemize}
    \setlength\itemsep{0em}
    \footnotesize{
            \item 100 training steps.
            \item 125 events per batch.
            \item A Learning rate of 0.01.
            \item Two hidden layers.
            \item 20 nodes per hidden layer.
            \item Exponential linear unit activation function for the nodes in the hidden layer.
            \item 15 output nodes (three output nodes corresponding to each Gaussian):
                \begin{itemize}
                \setlength\itemsep{0em}
                    \item 5 output nodes corresponding to $\mu_i$ - no activation function (identity operator).
                    \item 5 output nodes corresponding to un-normalised fractions $f_i$ - no activation function (identity operator).
                    \item 5 output nodes corresponding to $\sigma_i$, where the `activation' is the exponential function, to ensure that the widths of the Gaussians are positive.
                \end{itemize}
    }
    \end{itemize}

    The method is then as follows:
    \begin{enumerate}
        \item Train the NN to optimally separate signal and continuum. Save the weights, this is TF1.
        \item Create the ANN, and the classifying (the original) NN with the same architecture, and initialise the weights to that of the saved best model (that is the same architecture and weight values as TF1).
        \item For every (20,000 steps as there are four epochs and a batch size of 50) classifier training step and a given choice of $\lambda_{adv}$:
        \begin{enumerate}
            \item Train the ANN for the given number of adversary training steps (100 steps), where for each step:
                \begin{enumerate}
                    \item For every event in the batch (where the number of events in the batch is the adversary batch size, 125 events), get the $\mathit{NN}$ output from the classifier.
                    \item Using $\mathit{NN}$ and $\Delta E$ get the adversarial loss given by \ref{chapt8equation:adversaryLoss}.
                    \item Train the ANN given the adversarial loss, adversarial learning rate (0.01) and gradient descent algorithm of choice (Adam optimiser).
                \end{enumerate}
            \item Train the classifier neural network as normal for one training step, with the difference that the loss function is now given by \ref{chapt8equation:totalLoss} and has a dependence on $\Delta E$, as well as $\mathit{NN}$ and $\hat{y}$ (one or zero depending on if an event is signal or continuum).
        \end{enumerate}
        \item Save the weights of the classifying NN and use this updated NN for further analysis. These will be referred to as TF2.
    \end{enumerate}

    \begin{figure}[h!]
      \begin{center}
        \includegraphics[width=0.6 \columnwidth,height=!,angle=0]{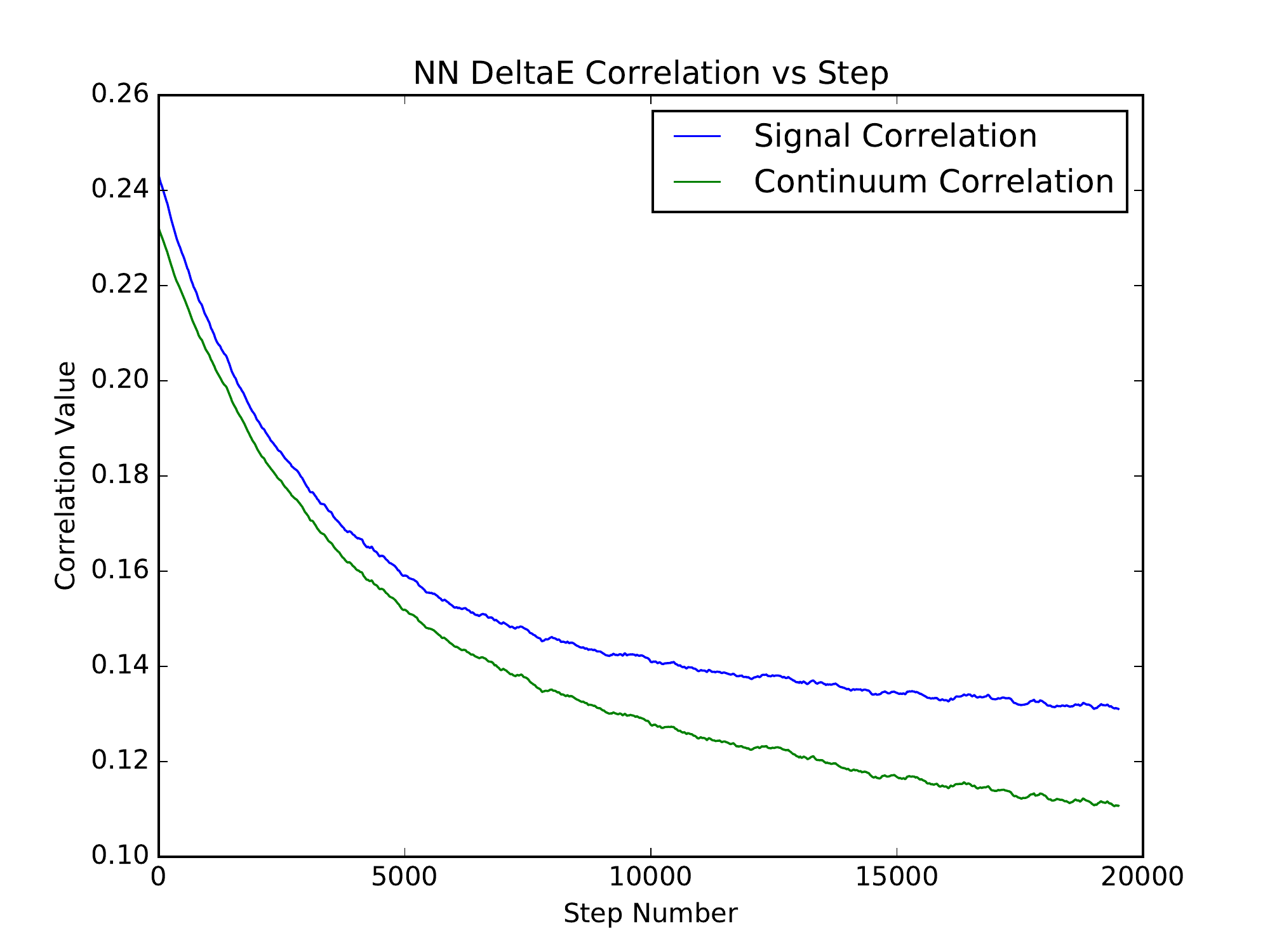}
      \end{center}
      \caption
      {Showing the signal (blue) and continuum (red) validation dataset correlations between $\Delta E$ and $\mathit{NN}$ as the training proceeds for TF2 and $\lambda_{adv} = 0.5$. This corresponds to 4 epochs, of 5000 classifier-training steps each, where the adversarial network is trained for 125 steps per classifier training step. Note that these correlations are in the validation data sets, and calculated over the entire range $0<\mathit{NN}<1$. }
      \label{figChapt8:CorrelationsVsStepNo}
    \end{figure}
    \begin{figure}[h!]
      \begin{center}
        \subfloat{
        \includegraphics[width=0.45 \columnwidth,height=!,angle=0]{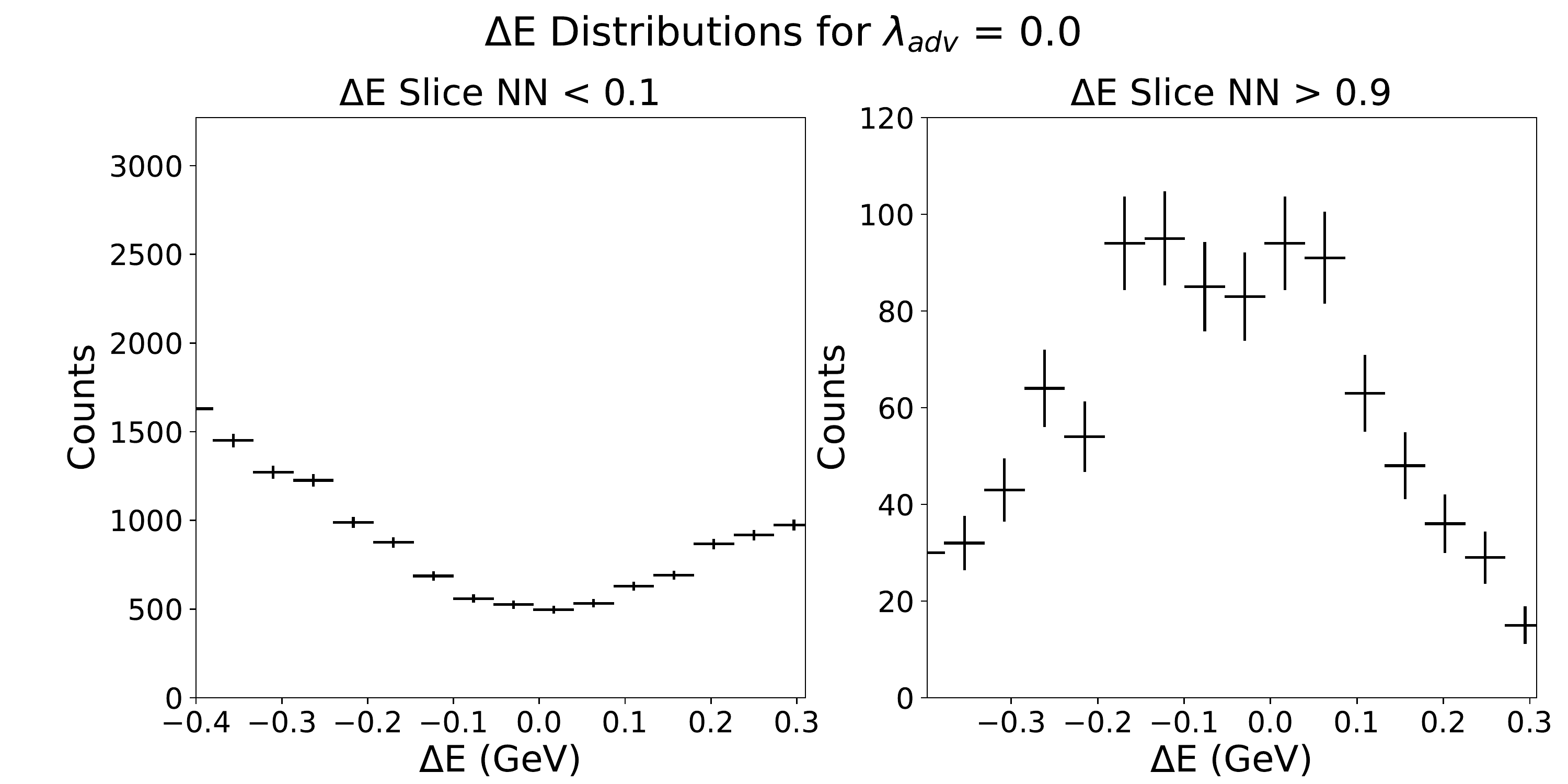}
        }
        \subfloat{
        \includegraphics[width=0.45 \columnwidth,height=!,angle=0]{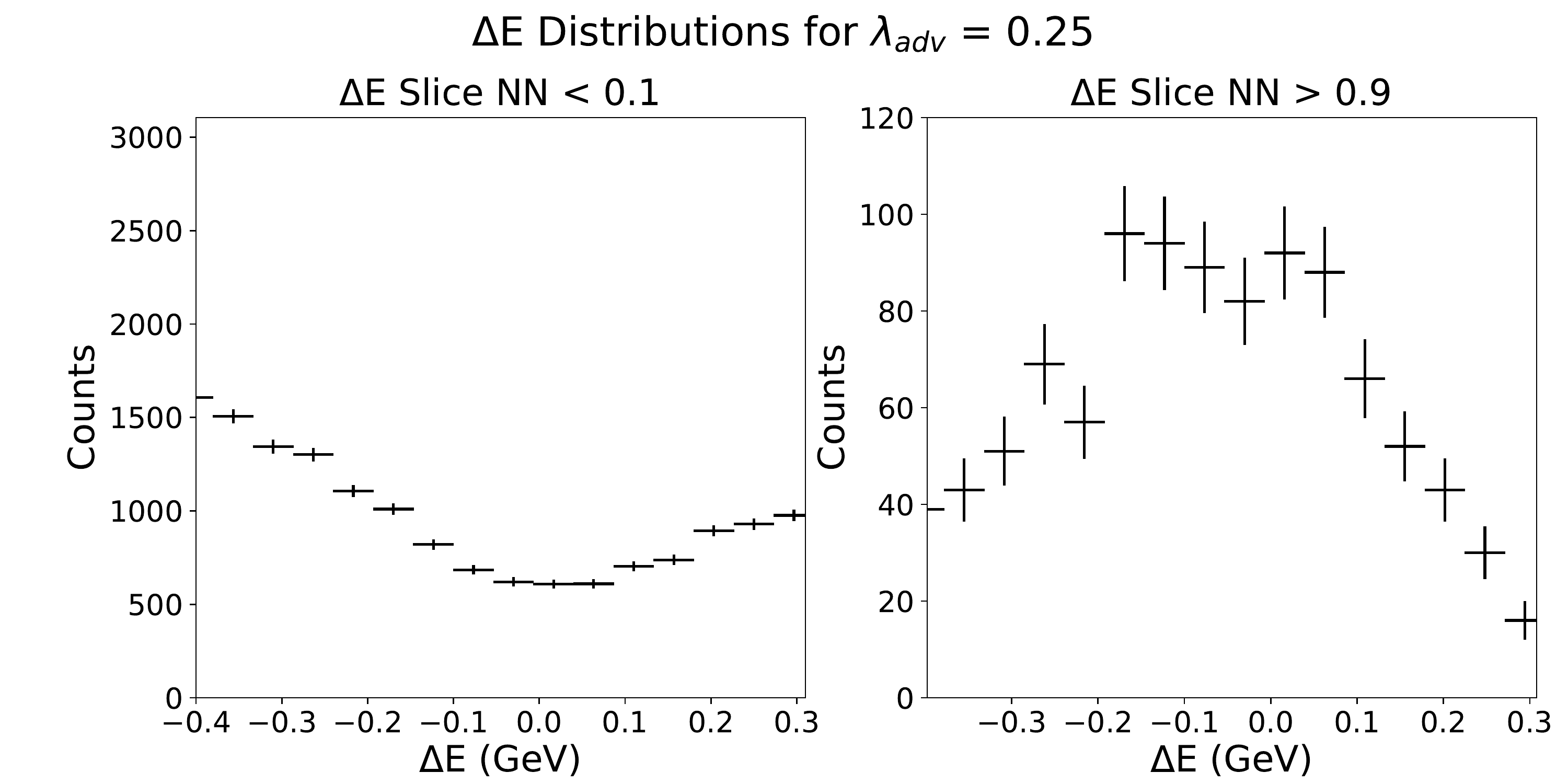}
        }\\
        \subfloat{
          \includegraphics[width=0.45 \columnwidth,height=!,angle=0]{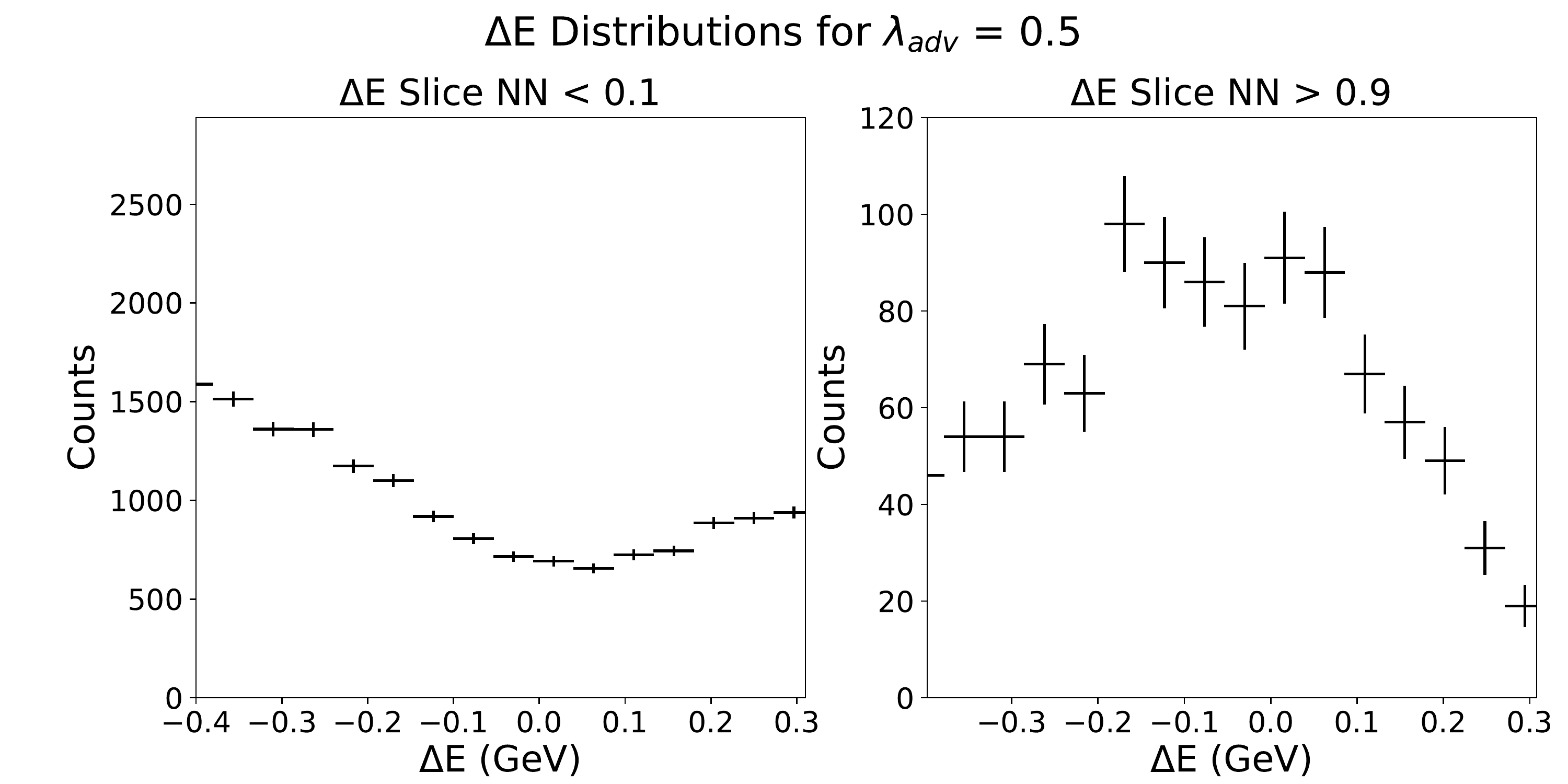}
        }
        \subfloat{
        \includegraphics[width=0.45 \columnwidth,height=!,angle=0]{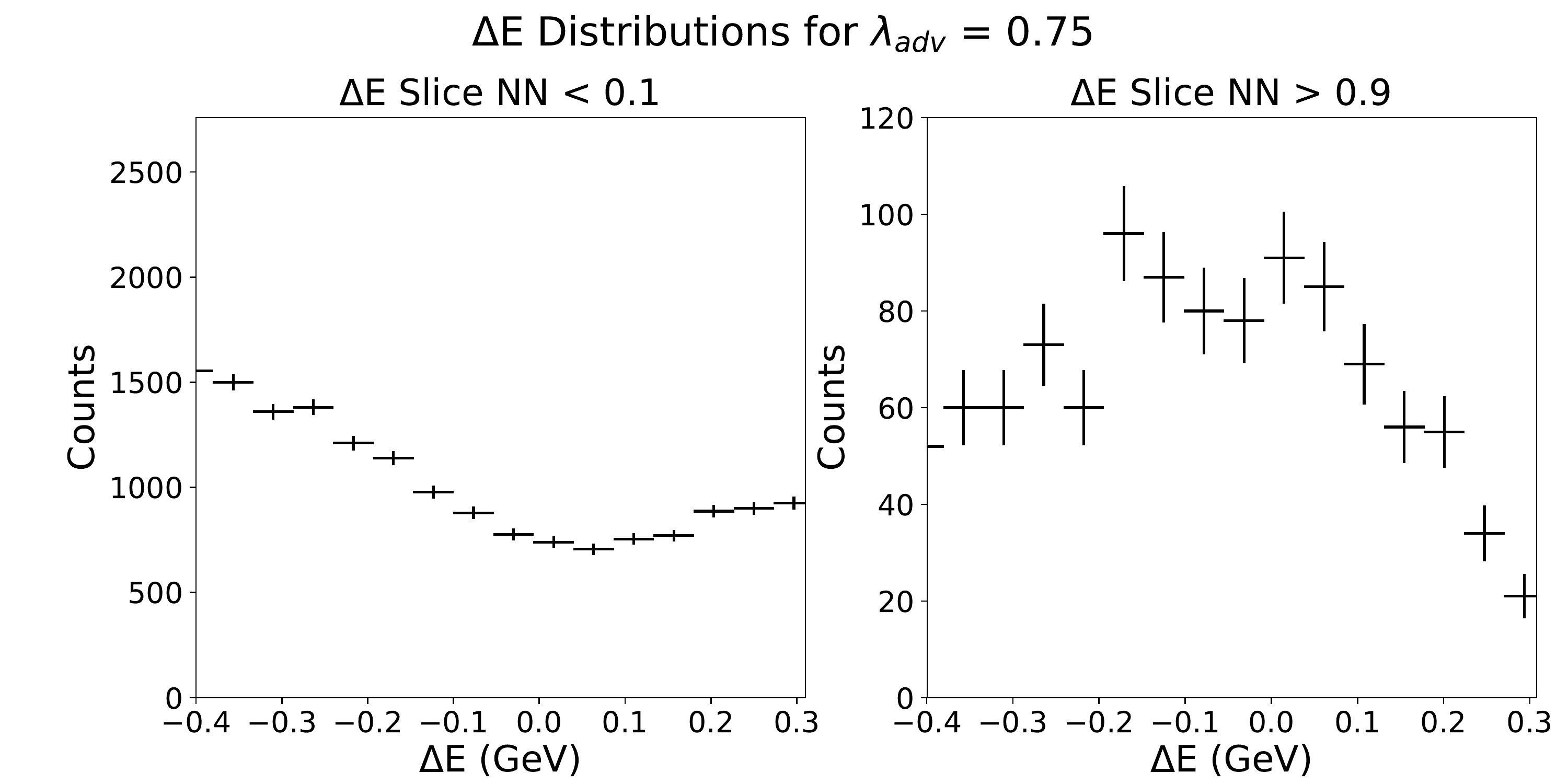}
        }\\
        \subfloat{
        \includegraphics[width=0.45 \columnwidth,height=!,angle=0]{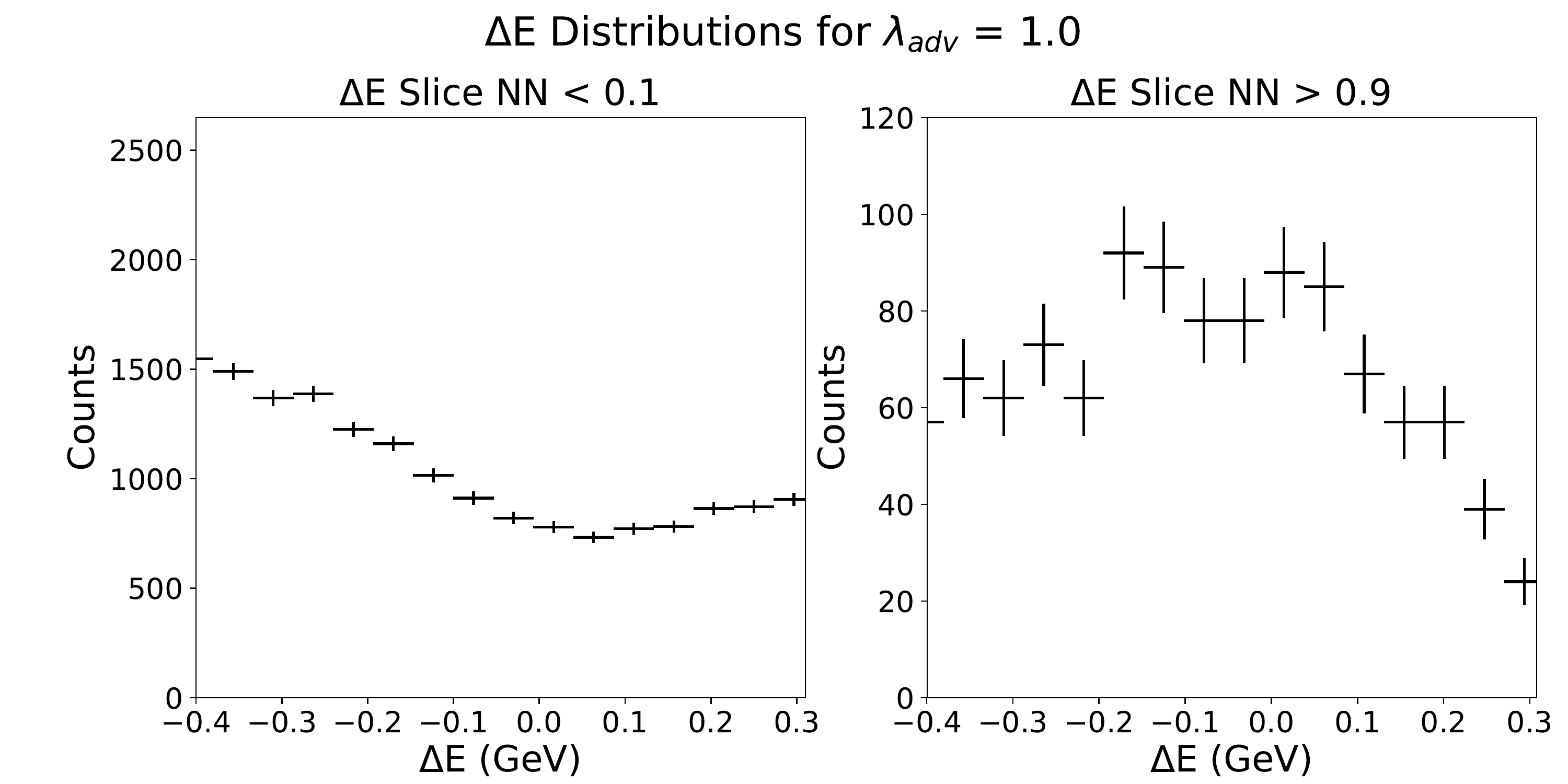}
        }
        \subfloat{
        \includegraphics[width=0.45 \columnwidth,height=!,angle=0]{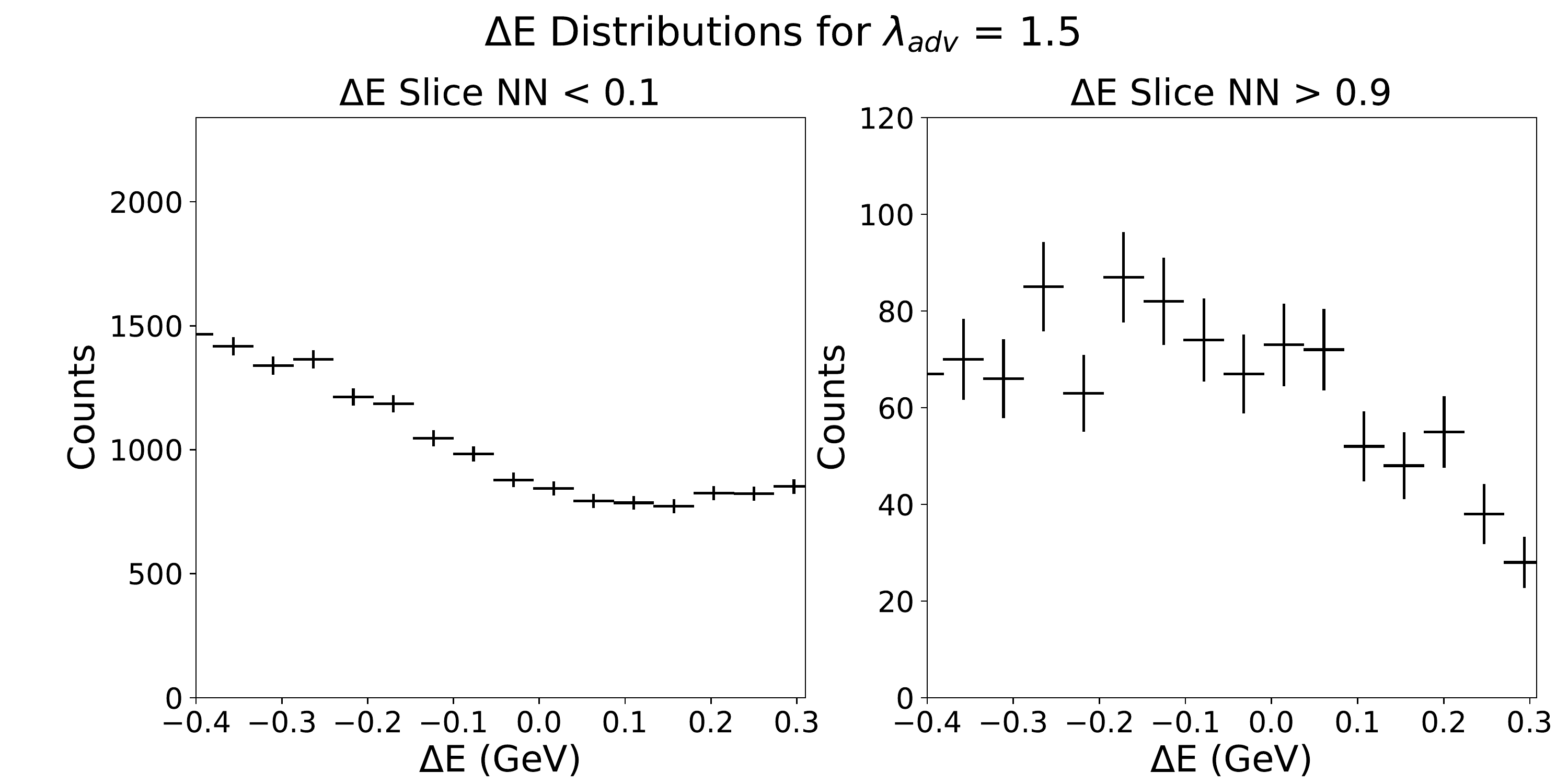}
        }\\
        \subfloat{
        \includegraphics[width=0.45 \columnwidth,height=!,angle=0]{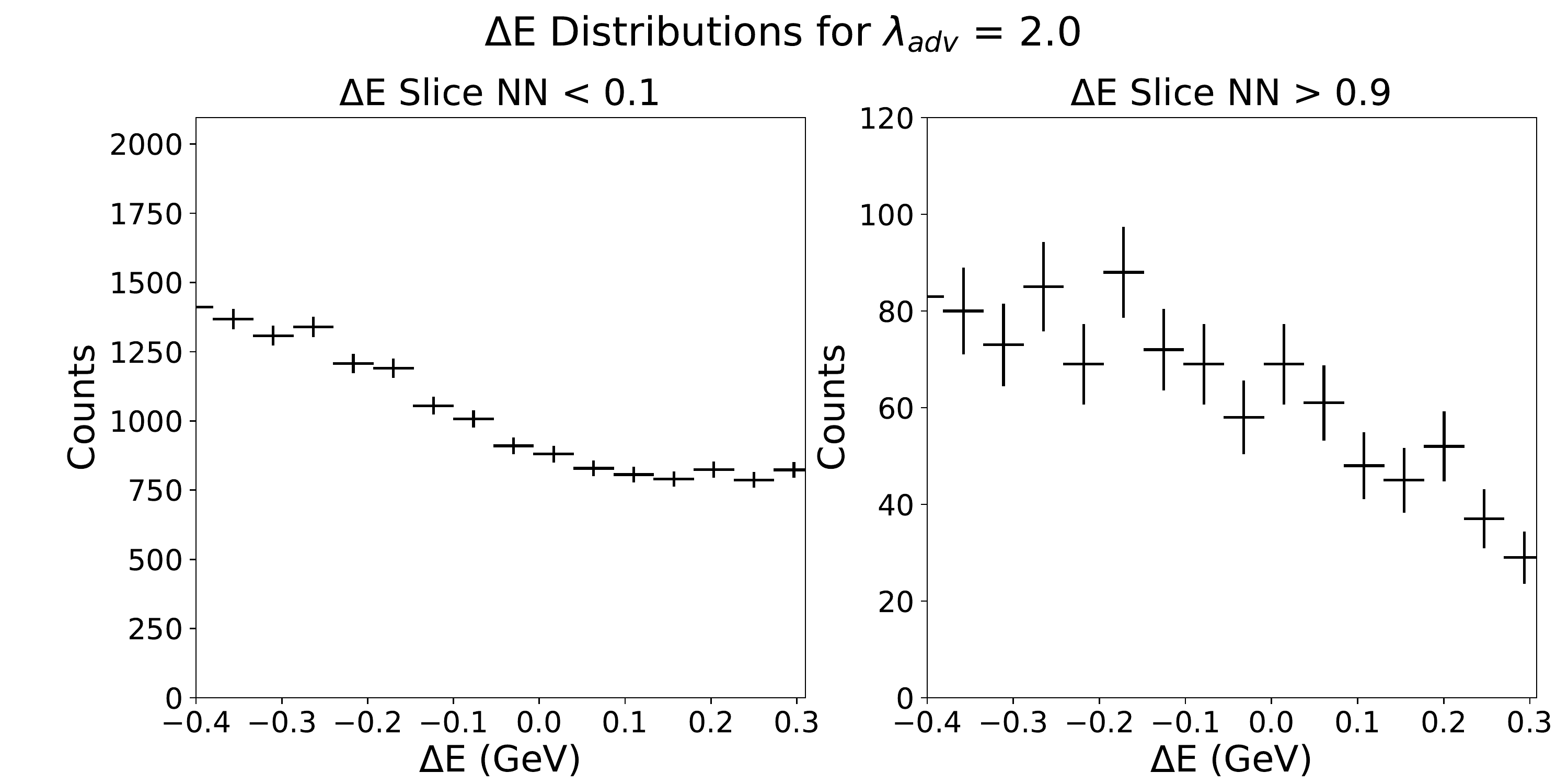}
        }
        \subfloat{
        \includegraphics[width=0.45 \columnwidth,height=!,angle=0]{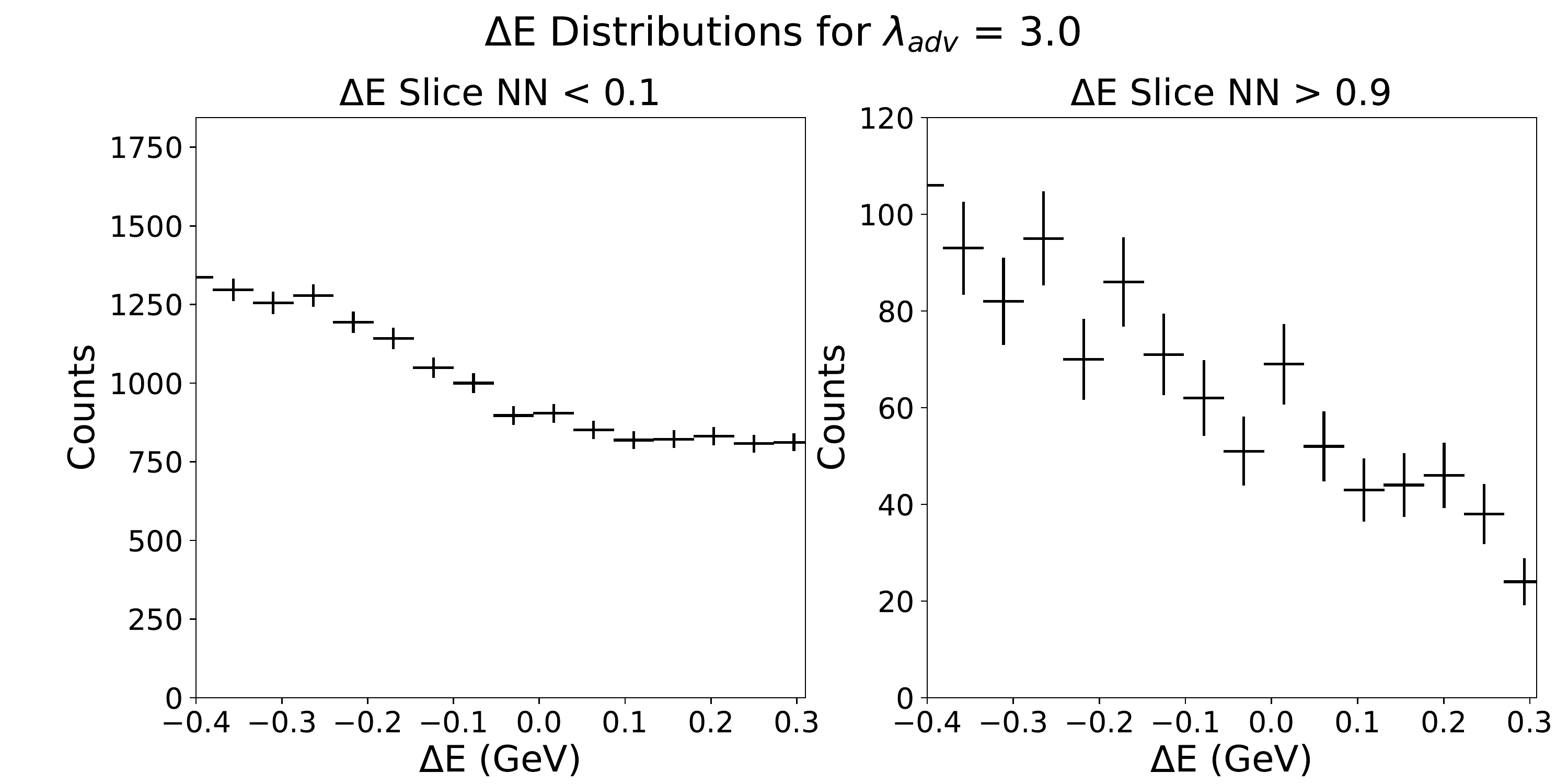}
        }\\
        \subfloat{
        \includegraphics[width=0.45 \columnwidth,height=!,angle=0]{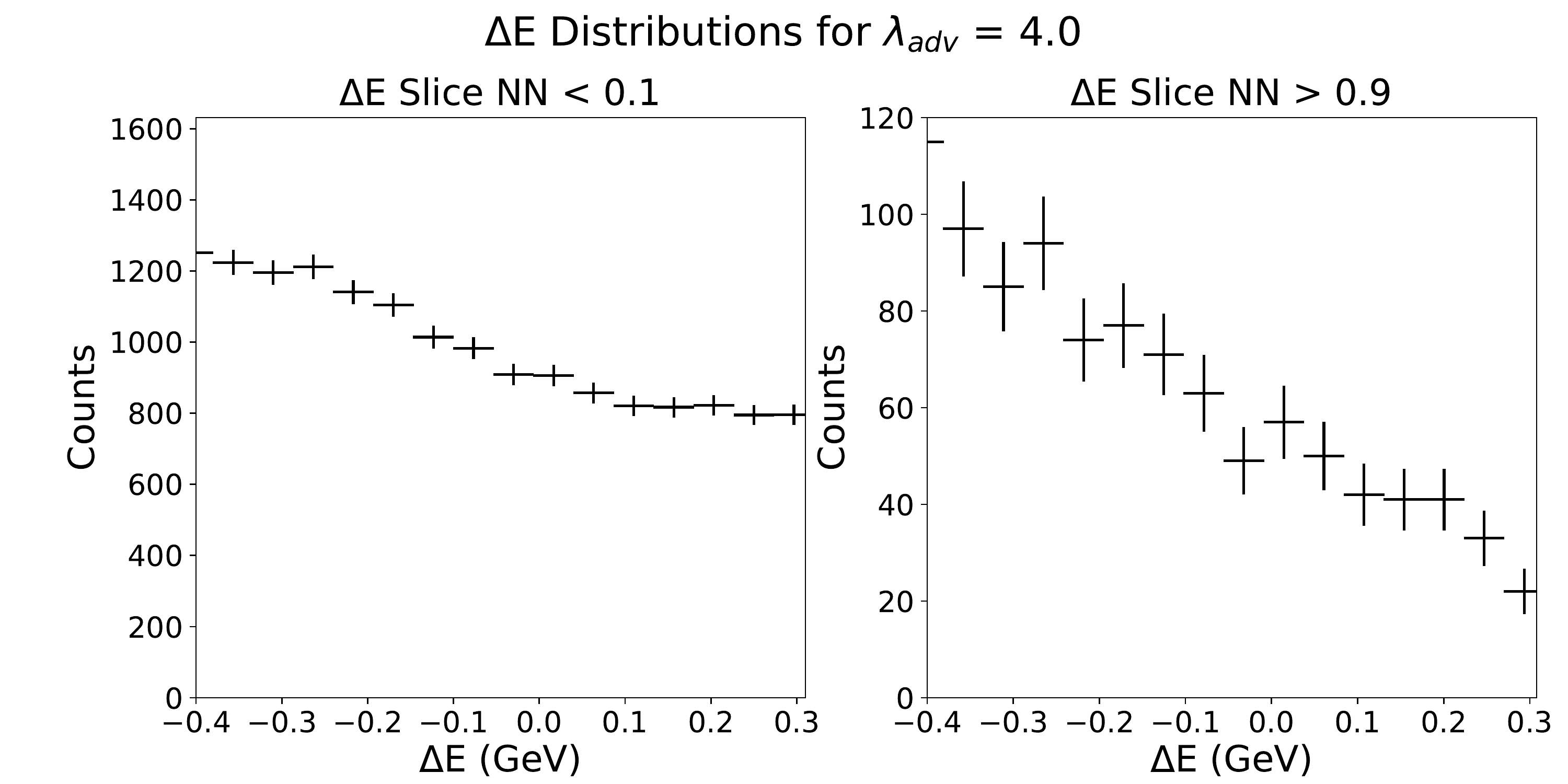}
        }
        \subfloat{
        \includegraphics[width=0.45 \columnwidth,height=!,angle=0]{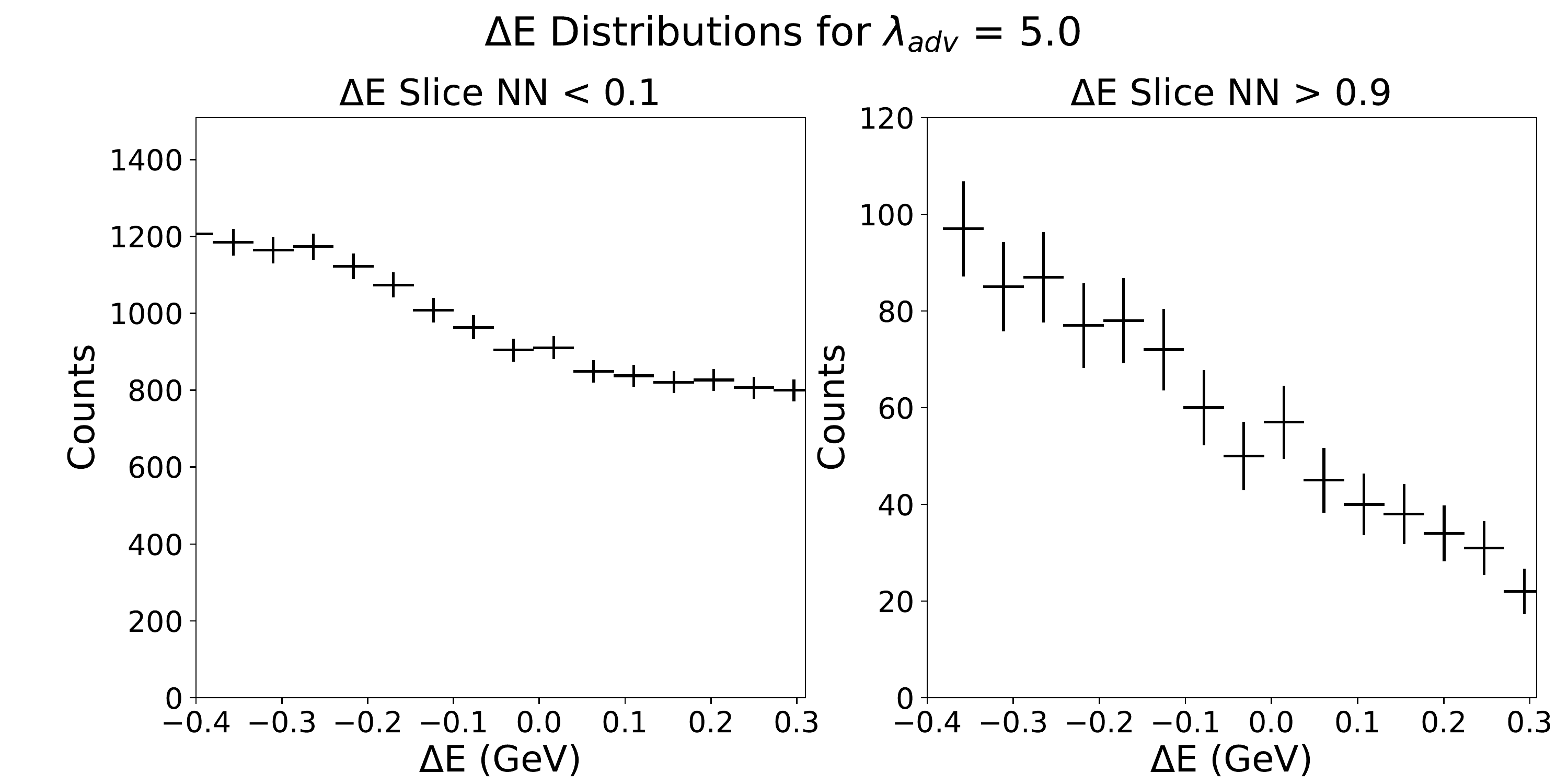}
        }
      \end{center}
      \caption
      {Showing the continuum $\Delta E$ slices at $\mathit{NN}<0.1$ and $\mathit{NN}>0.9$ for $\lambda_{adv}$ between 0.0 and 5.0 as shown.}
      \label{figChapt8:DeleNNSlicesLambdaAdv}
    \end{figure}

    \subsection{Analysis of the TensorFlow With Adversary Neural-Network Performance}

    An investigation was performed by repeating the procedure for $\lambda_{adv}$ values of 0.25, 0.5, 0.75, 1.0, 1.5, 2.0, 3.0, 4.0 and 5.0.
The testing data sets are processed by each of the NNs (TF2) trained with the different $\lambda_{adv}$ values. The continuum rejection rates in the full ($-0.4<\Delta E<0.3$) and tight ($-0.1<\Delta E<0.1$) ranges, for each of these NNs along with the NeuroBayes NN, BDT and original TensorFlow NN without adversary, are shown in Table \ref{chapt8:advLambdaNetResults}. As before, these rejection rates are calculated after imposing selection criteria on $\mathit{NN}$ such that $92.5\%$ of signal remains in the full $\Delta E$ range.
    
    We find that the $\Delta E - \mathit{NN}$ sculpting steadily decreases with increasing $\lambda_{adv}$. The $\Delta E- \mathit{NN}$ correlation decreases until $\lambda_{adv} = 1.0$, after this, the correlation becomes increasingly negative even as the scultping effect continues to diminish. The background rejection rate also becomes worse for  $\lambda_{adv} > 1.0$. During the training process, the $\Delta E- \mathit{NN}$ correlation decreases with step number at small$\lambda_{adv}$, while for larger values of $\lambda_{adv}$, the correlations ``bounce'' before dminishing again.  This behaviour can be explained by the competing NNs. For smaller $\lambda_{adv}$ values, they settle to a coupled situation where training in one network is counteracted by the other. The larger $\lambda_{adv}$ values see the adversarial network dominate quickly before the classifier has had enough training steps to counteract it. The $\Delta E - \mathit{NN}$ correlations against the number of training steps for $\lambda_{adv}=0.5$ is shown in Figure \ref{figChapt8:CorrelationsVsStepNo}. The continuum MC (testing dataset) $\Delta E$ distributions (at different $\mathit{NN}$ slices) for each $\lambda_{adv}$ are shown in Figure \ref{figChapt8:DeleNNSlicesLambdaAdv}. We choose  $\lambda_{adv} = 1.5$ as the point of comparison for the rest of the discussion as this is best compromise between minimizing the scuplting, the correlation with $\Delta E$ while retaining the best background rejection rates.

        \begin{table}[ht]
            \begin{center}
            \begin{tabular}{ | c | c | c | c |}
            \hline
             $\lambda_{adv}$ & RR Full-$\Delta E$ Range & RR Tight-$\Delta E$ Range & Correlation with $\Delta E$\\
            \hline
            N/A (TF1) & $81.3$\% & $63.2$\% & $0.114$\\
            \hline
            0.25 (TF2)& $80.6$\% & $65.1$\% & $0.080$\\
            \hline
            0.50 (TF2)& $79.7$\% & $66.6$\% & $0.057$\\
            \hline
            0.75 (TF2)& $80.0$\% & $67.6$\% & $0.025$\\
            \hline
            1.00 (TF2)& $78.4$\% & $67.7$\% & $0.012$\\
            \hline
            1.50 (TF2)& $74.6$\% & $67.4$\% & $-0.024$\\
            \hline
            2.0 (TF2)& $70.2$\% & $66.6$\% & $-0.057$\\
            \hline
            3.0 (TF2)& $64.3$\% & $63.0$\% & $-0.106$\\
            \hline
            4.0 (TF2)& $57.1$\% & $57.6$\% & $-0.145$\\
            \hline
            5.0 (TF2)& $51.9$\% & $53.4$\% & $-0.175$\\
            \hline
            (NB) & $66.7$\% & $64.08$\%& $0.058$\\
            \hline
            (NB, reduced) & $63.0$\% & $64.08$\%& $-0.001$\\
            \hline
            (BDT) & $73.7$\% & $67.2$\% & $0.262$\\
            \hline
            (BDT, reduced) & $66.9$\% & $66.5$\% & $0.054$ \\
            \hline
            \end{tabular}
            \end{center}
            \caption{The continuum MC rejection rates (RR) for the full and tight $\Delta E$ regions with a signal acceptance of $92.5\%$. Also shown in column 4 is the final correlation between  $\Delta E$ and the discriminating variable for the continuum MC background. `N/A (TF1)' refers the the TensorFlow NN trained without adversary, `(TF2)' refer to the adversarialy trained NNs, `(NB)' refers to the NeuroBayes NN, '(NB, reduced)' refers to NeuroBayes with the $R^{so}_{20}$, $R^{oo}_{0}$, $R^{oo}_{2}$ and $R^{so}_{22}$ variables removed. '(BDT)' is the Boosted Decision Tree algorithim. '(BDT, reduced)' is the BDT with the $R^{so}_{20}$, $R^{oo}_{0}$, $R^{oo}_{2}$ and $R^{so}_{22}$ variables removed.}
        \label{chapt8:advLambdaNetResults}
        \end{table}

\section{Validation With Off-Resonance}
\label{sec6}
\begin{figure}[h!]
          \begin{center}
            \includegraphics[width=0.95 \columnwidth,height=!,angle=0]{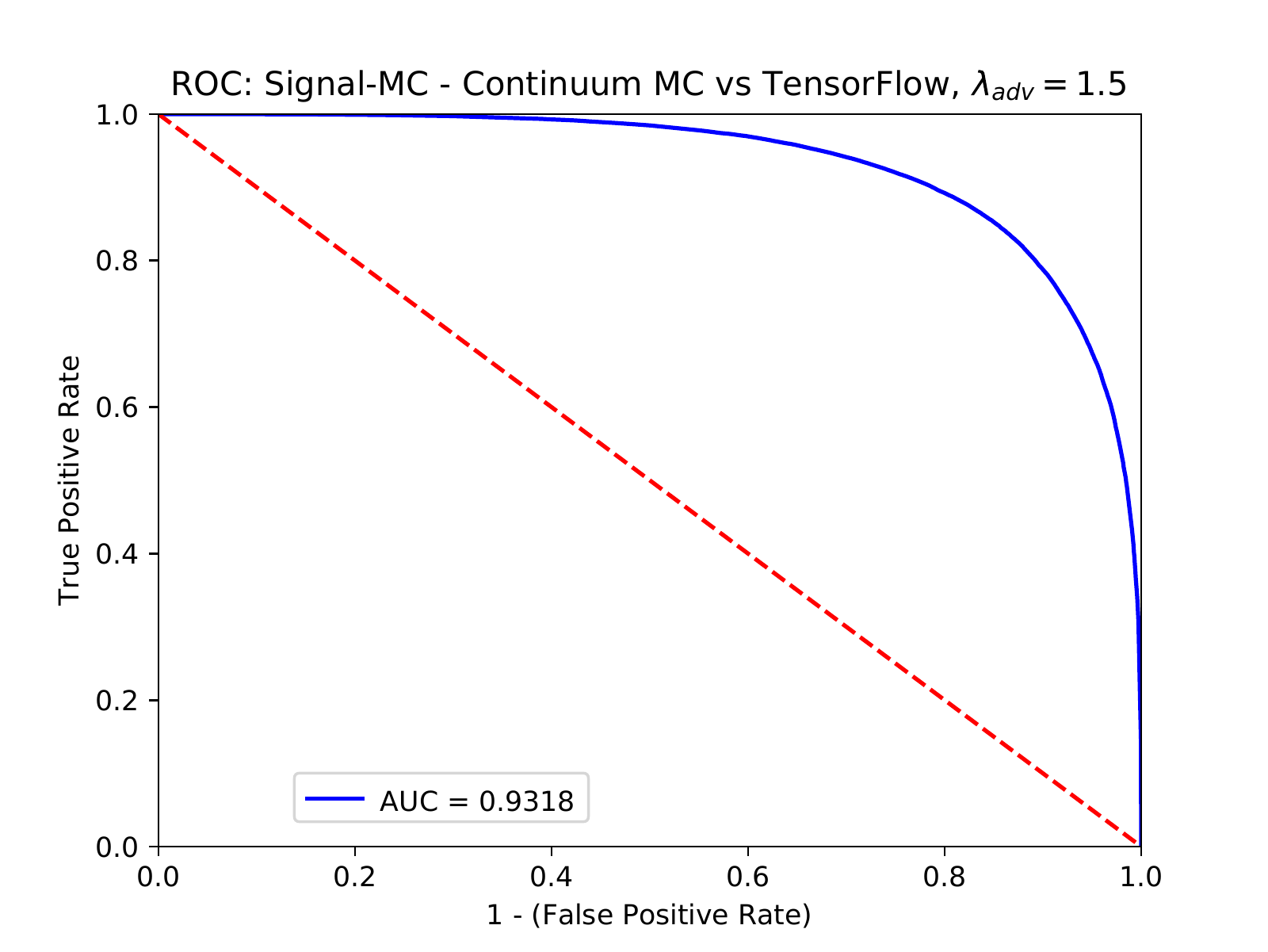}
          \end{center}
          \caption
          {Showing the TF2 ROC curve for signal and continuum MC for $\lambda_{adv}=1.5$. }
          \label{figChapt8:TFNNSigContEqualNumbers}
        \end{figure}

\begin{figure}[h!]
          \begin{center}
            \includegraphics[width=0.95 \columnwidth,height=!,angle=0]{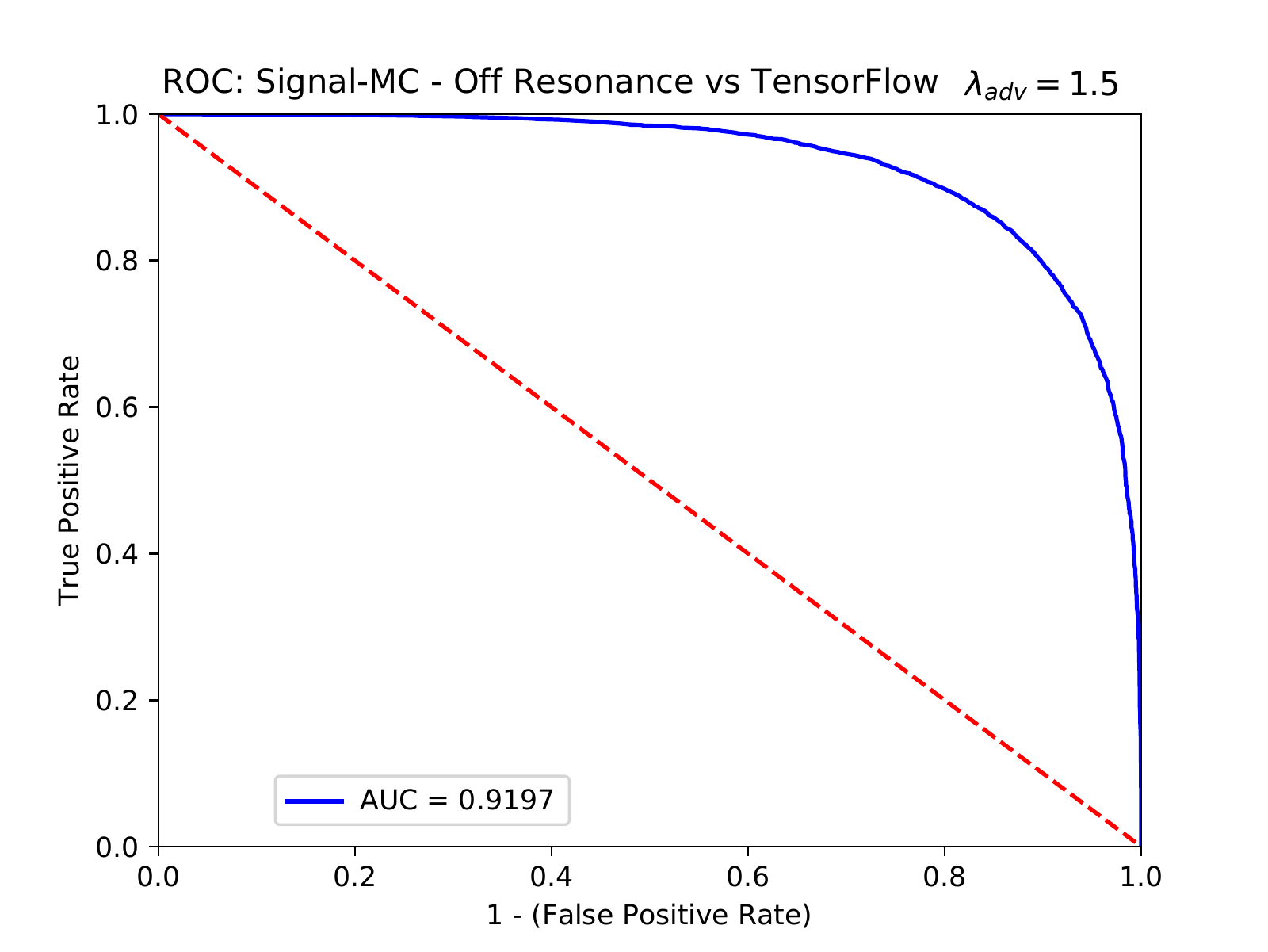}
          \end{center}
          \caption
          {Showing the TF2 ROC curve for signal and off-resonance for $\lambda_{adv}=1.5$, where the noisiness is due to the smaller sample size of the off-resonace data. }
          \label{figChapt8:TFNNSigOffResEqualNumbers}
        \end{figure}

    To fully evaluate the performance of the TF2, we process real off-resonance data to test the MC simulations. The performance of TF2 with $\lambda_{adv}=1.5$ for continuum MC and off-resonance are shown in Figures \ref{figChapt8:TFNNSigContEqualNumbers} and \ref{figChapt8:TFNNSigOffResEqualNumbers} respectively.

        \begin{figure}[h!]
          \begin{center}
            \subfloat{
            \includegraphics[width=0.95 \columnwidth,height=!,angle=0]{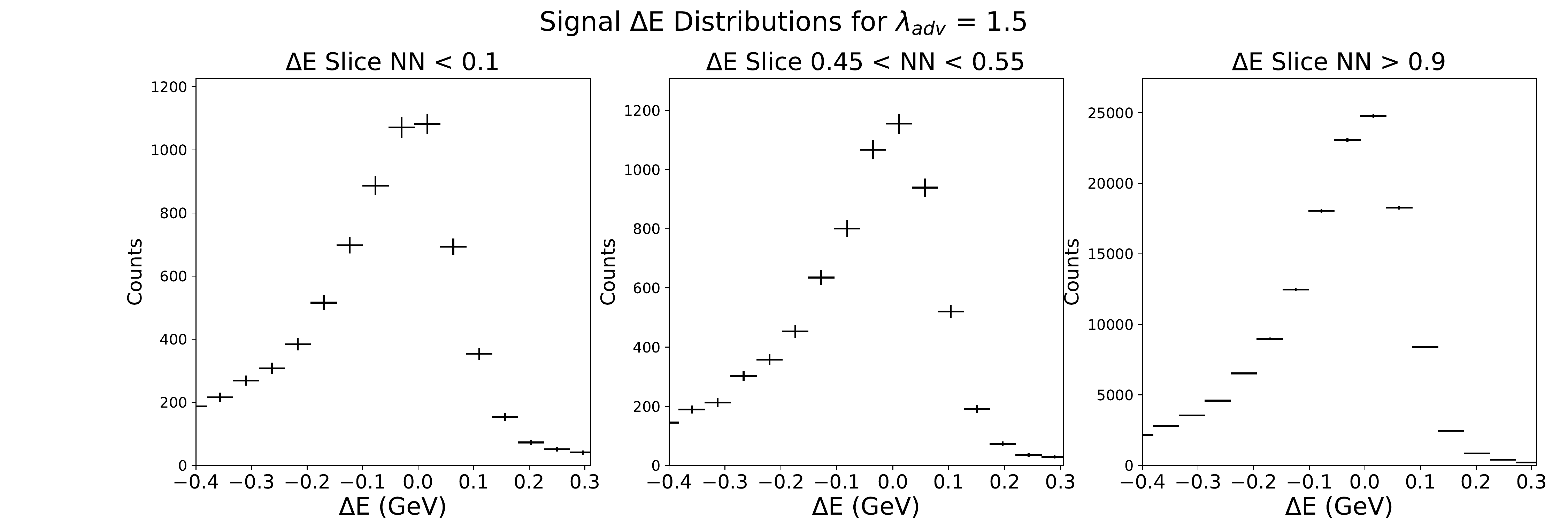}
            } \\

            \subfloat{
            \includegraphics[width=0.95 \columnwidth,height=!,angle=0]{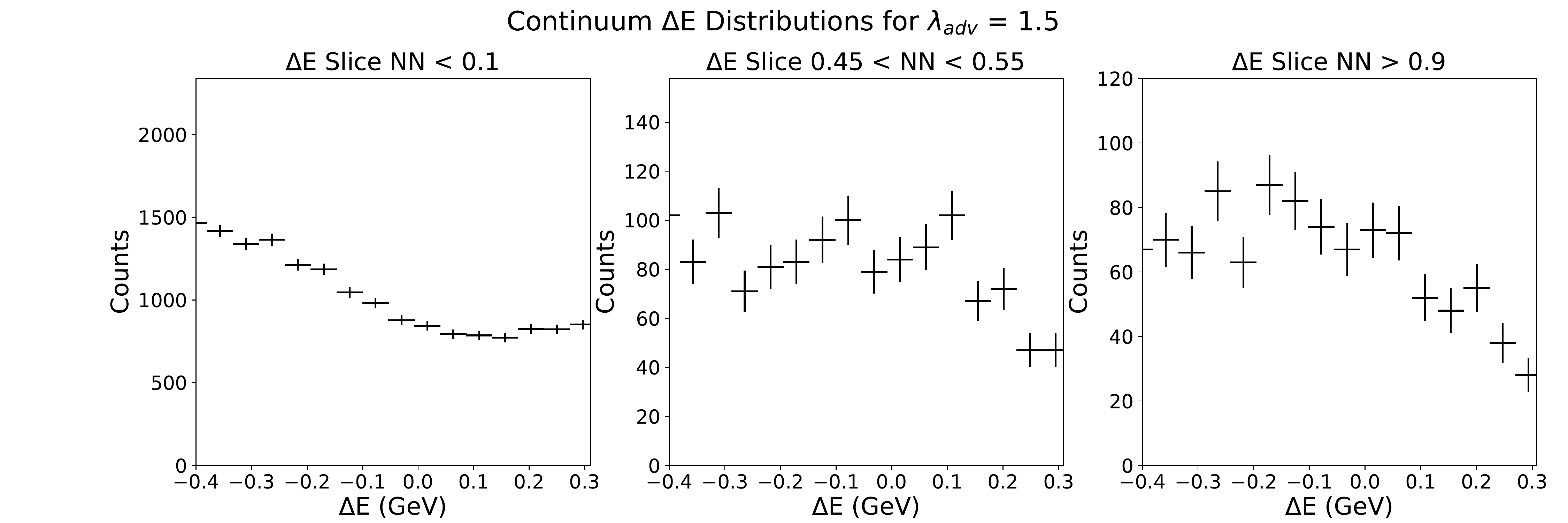}
            } \\
            \subfloat{
            \includegraphics[width=0.95 \columnwidth,height=!,angle=0]{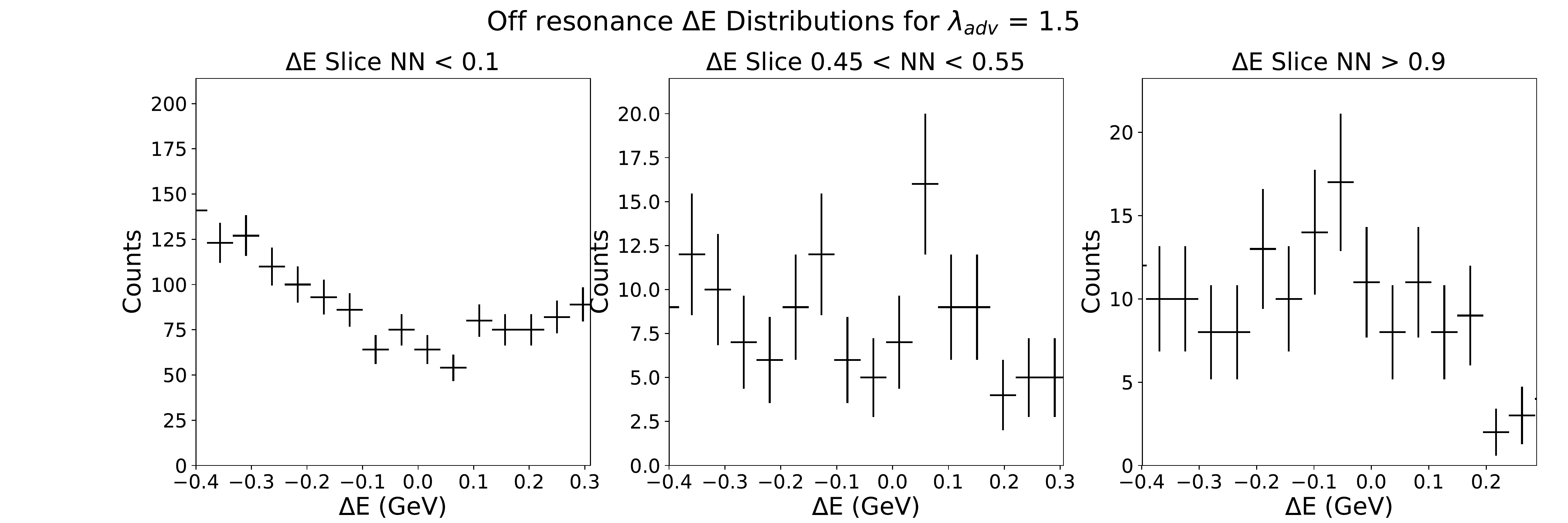}
            } \\
          \end{center}
          \caption
          {Showing the signal MC (top row), continuum MC (middle row) and off-resonance data (bottom row) $\Delta E$ distributions for $\mathit{NN}<0.1$ (left column), $0.45<\mathit{NN}<0.55$ (middle column) and $0.9<\mathit{NN}$ (right column) for TF2 with $\lambda_{adv}=1.5$. }
          \label{figChapt8:DeleNNSlicesSigContOffRes}
        \end{figure}
        
        The $\Delta E$ distributions at different $\mathit{NN}$ slices for signal MC, continuum MC and off-resonance data are shown in Figure  \ref{figChapt8:DeleNNSlicesSigContOffRes}.
        
        Finally, the off-resonance rejection rates (for $\mathit{NN}_{cut}$ keeping $92.5\%$ signal) for the NeuroBayes NN, TF1 and TF2 with $\lambda_{adv}=1.5$ are shown in table \ref{chapt8:advLambdaNetResultsWITHOFFRES}. As compared to the results in Table \ref{chapt8:advLambdaNetResults}, the rejection rates are similar although poorer for off-resonance (to be expected as the NNs were trained entirely with MC simulations which employ parameterized theoretical model to simulate $e^{+}e^{-}\to q\bar{q}$ interactions) but are consistently worse regardless of which classifier was used. Moreover, the trends identified earlier for TF1 and TF2 versus NB are the same. TF1 provides the best rejection over the full $\Delta E$ range and TF2 provides the best performance in the tight $\Delta E$ range. We conclude that the NN's do significantly reduce the experimental continuum background at rates close to those predicted by the MC studies.

        \begin{table}[ht]
            \begin{center}
            \begin{tabular}{ | c | c | c | }
            \hline
             $\lambda_{adv}$ & Off-Resonance RR Full-$\Delta E$ Range & Off-Resonance RR Tight-$\Delta E$ Range\\
            \hline
            N/A (TF1) & $77.7$\% & $58.6$\%\\
            \hline
            1.50 (TF2)& $69.0$\% & $60.5$\%\\
            \hline
            N/A (NB) & $61.1$\% & $59.8$\%\\
            \hline
            N/A (BDT) & $68.7$\% & $60.4$\%\\
            \hline
            \end{tabular}
            \end{center}
            \caption{The off-resonance rejection rates (RR) for the full and tight $\Delta E$ regions. The signal acceptance was set to $92.5\%$}
        \label{chapt8:advLambdaNetResultsWITHOFFRES}
        \end{table}


\section{Conclusions}
\label{sec7}
While the purpose-built deep NN developed with TensorFlow (TF1) has better continuum suppression than the commercial NeuroBayes (NB) package and the Boosted Descision Tree (BDT) algorithim, it achieved this at the cost of significant sculpting of the $\Delta E$ distribution which is used to distinguish signal from background. In fact in the most sensitive region to signal, -0.1 GeV $< \Delta E <$ 0.1 GeV, TF1 had a poorer performance than NB and BDT. Setting the classifier to accept $92.5\%$ of the signal we find background rejection rates of 63.4\% vs 64.8\% and 67.2\% for TF1, NB and BDT respectively. We then employed an adversarial NN (TF2) to counter-act this correlation. With $\lambda_{adv}=1.5$, TF2 achieved improved performance compared to NB in both the most sensitive (67.4\% vs 64.1\%) and full (74.6\% vs 66.7\%) regions of $\Delta E$. While the background rejection rate of TF2 with $\lambda_{adv}=1.5$ was only slightly better than the BDT in the most sensitive region (67.4\% vs 67.2\%), the correlation with $\Delta$E for TF2 was significantly better and the BDT (-0.024 vs 0.264). While some sculpting remains at  $\lambda_{adv}=1.5$, this $\Delta E$ distribution is still significantly different from that of the signal and the overall correlation between TF2 and $\Delta E$ is close to minimized at $\lambda_{adv}=1.5$. Consequently the adversarial neural network allows us to employ discriminating variables correlated with $\Delta E$ to improve background rejection.

Rogozhnikov et al. \cite{cross_ent_boost} suggest an alternative approach to reduce the effect of scuplting. This is to design a cross entropy which explicitly penalizes correlations with non-classification variables. This would be worth investigating in a future work.

\section{Acknowledgements}
We gratefully acknowledge our colleagues within the Belle collaboration for all the work required to make this investigation possible and for their permission to employ Belle data and tools for this study. We would particularly like to thank Thomas Keck of the Karlsruher Institut f{\"u}r Technologie for very helpful discussions about adversarial neural networks. We thank the KEKB group for the excellent operation of the accelerator; the KEK cryogenics group for the efficient operation of the solenoid; and the KEK computer group, the National Institute of Informatics, and the PNNL/EMSL computing group for valuable computing and SINET5 network support. We acknowledge financial support from the Australian Research Council.



\bibliographystyle{elsarticle-num}

\bibliography{references}

\end{document}